\documentclass[prd,aps,showpacs,nofootinbib,tightenlines]{revtex4}
\usepackage{mathrsfs}
\usepackage{amsmath}
\usepackage{amssymb}
\usepackage{epsfig}
\usepackage{graphicx}
\usepackage{booktabs}
\usepackage{multirow}
\usepackage{subfigure}
\usepackage{color}
\usepackage{slashed}

\begin{document}
\newcommand{\psl}{ p \hspace{-1.8truemm}/ }
\newcommand{\nsl}{ n \hspace{-2.2truemm}/ }
\newcommand{\vsl}{ v \hspace{-2.2truemm}/ }
\newcommand{\epsl}{\epsilon \hspace{-1.8truemm}/\, }
\title{
$CP$-violating observables in four-body $B\rightarrow \phi(\rightarrow K\bar K)K^*(\rightarrow K\pi)$ decays}
\author{Chao-Qi Zhang$^1$}
\author{Jia-Ming Li$^1$}
\author{Meng-Kun Jia$^1$}
\author{Ya Li$^2$} \email [Corresponding author: ] {liyakelly@163.com}
\author{Zhou Rui$^1$}\email [Corresponding author: ]{jindui1127@126.com}
\affiliation{$^1$ College of Sciences, North China University of Science and Technology, Tangshan, Hebei 063210, China}
\affiliation{$^2$ Department of Physics, College of Science, Nanjing Agricultural University, Nanjing, Jiangsu 210095, China}
\date{\today}

\begin{abstract}
We analyze the four-body $B\rightarrow \phi(\rightarrow K\bar K)K^*(\rightarrow K\pi)$ decays in the perturbative QCD approach,
where the invariant mass of the $K\bar K$ ($K\pi$) system is limited in a window of
$\pm 15$ MeV ($\pm150$ MeV) around the nominal $\phi(K^*(892))$ mass.
In addition to the dominant $P$-wave resonances, two important $S$-wave backgrounds
in the selected invariant mass region are also accounted for.
Angular momentum conservation allows six helicity amplitudes  to contribute,
including three $P$ waves, two single $S$ waves, and one double $S$ wave, in the decays under study.
We calculated the branching ratio for each component and found sizable $S$-wave contributions,
which  coincide roughly with the experimental observation.
The obtained branching ratios of $B^{0(+)}\rightarrow \phi K^{*0(+)}$ are comparable
with the previous theoretical predictions and support the experimental measurements,
whereas the predicted $\mathcal{B}(B^0_s\rightarrow \phi \bar K^{*0})$ is an order of magnitude  smaller than
the current world average in its central value.
The longitudinal polarizations are predicted to be around 0.7,
consistent with previous PQCD results but larger than the  world average values.
Aside from the direct $CP$ asymmetries, the  true and fake triple product asymmetries,
originating from the interference between the perpendicular polarization amplitude and other helicity amplitudes,
are also calculated in this work.
In the special case of the neutral modes, both the direct $CP$ asymmetries and true triple product asymmetries
are expected to be zero due to the vanishing weak phase difference.
The direct $CP$ asymmetries for the $B^+$ mode are predicted to be tiny, of order $10^{-2}$,
since the tree contributions are suppressed strongly with respect to the penguin ones.
The true triple product asymmetries   have shown no significant deviations from zero.
In contrast, large fake asymmetries are observed in these decays,
indicating the presence of significant final-state interactions.
We give the theoretical predictions of the $S$-wave induced triple product asymmetries for the first time,
which is consistent with current LHCb data and would be
checked  with future measurements from Belle and $BABAR$ experiments
if the $S$-wave components can be properly taken into account in the angular analysis.

\end{abstract}


\maketitle

\section{Introduction}
The phenomenology of $B$ decays to two light vector mesons
 provides unique opportunities for understanding the mechanism of hadronic weak decays
and their $CP$ asymmetry, and probing the new physics (NP) beyond the standard model (SM).
Angular momentum conservation leads to three independent configurations of
the vector mesons which reflects into three amplitudes.
In the transversity basis,
the decay amplitude can be  decomposed into three independent components
$A_{0}$, $A_{\parallel}$, and $A_{\perp}$~\cite{Sinha:1997zu,Kramer:1991xw},
which correspond to longitudinal, parallel, and perpendicular polarizations of the final-state spin vectors, respectively.
Experimentally they are at least four-body decays~\cite{Belle-II:2018jsg},
since a vector resonance is usually detected via its decay $V\rightarrow PP'$ with $P^{(')}$ being a pseudoscalar.
As the vector meson has a relatively broad width,
there is generally a background due to the (resonant or nonresonant) scalar production
of the two pseudoscalars~\cite{Stone:2008ak,Xie:2009fs} around the vector resonance region.
In this case, it is necessary to add another three scalar amplitudes to the angular analysis in the presence of the scalar background~\cite{Bhattacharya:2013sga}.
Then one can extract more $CP$-violating observables  from the interference of the various helicity amplitudes.
Therefore, the four-body charmless $B$ decays  through two vector intermediate states
are rich in $CP$-violating phenomena in the flavor sector involving quarks.

A four-body decay gives rise to three independent final momenta $\vec{p}_i$ with $i=1,2,3$
in the rest frame of the decaying parent particle, and allows one to form a scalar triple product (TP)
of  $\vec{p}_1\cdot (\vec{p}_2 \times \vec{p}_3)$.
Obviously, it is odd under both parity and time reversal, and thus constitutes a potential signal of  $CP$ violation assuming $CPT$ invariance.
One can compare event distributions for positive TP against those with negative TP to construct a triple product asymmetry (TPA),
\begin{eqnarray}\label{eq:at}
A_T\equiv \frac{\Gamma(TP>0)-\Gamma(TP<0)}{\Gamma(TP>0)+\Gamma(TP<0)},
\end{eqnarray}
where $\Gamma$ is the partial decay rate in the indicated TP range.
However, since both the final-state interaction and  $CP$ violation can produce the nonzero TPAs,
one has to compare this asymmetry with a corresponding quantity in the $CP$ conjugate process to obtain the "true" $CP$ violation signal.
Furthermore, unlike the direct $CP$ violation, a nonzero true TPA
does not require the presence of a nonzero strong phase~\cite{Bensalem:2000hq},
while it is maximal when the strong phase difference vanishes.
In this case,
it could be more promising to search for TPA than direct $CP$ asymmetries in $B$ decays.
Therefore, the TPA is one powerful tool for displaying $CP$ violation in weak four-body decays~\cite{Datta:2003mj,Gronau:2011cf}.
Further information on this rich subject may be found in Refs.~\cite{Bensalem:2000hq,Datta:2003mj,Datta:2011qz,Gronau:2011cf,Datta:2012ky,Bhattacharya:2013sga,Patra:2013xua,Durieux:2015zwa,Bevan:2014nva}.

Four-body decays of the $B$ meson are more complicated than the
two-body case, specifically where both nonresonant and resonant contributions exist.
In our previous works~\cite{zjhep,Li:2021qiw}, the PQCD factorization formalism
based on the quasi-two-body decay mechanism~\cite{Wang:2015uea,Krankl:2015fha,Boito:2017jav}
for four-body $B$ meson decays has been well established.
That is, a four-particle final state is obtained through two intermediate resonances.
The resonances decaying into the meson pair are modeled by nonperturbative two-hadron distribution amplitudes (DAs)~\cite{G,G1,DM,Diehl:1998dk,Diehl:1998dk1,Diehl:1998dk2,MP},
which collect both resonant and nonresonant contributions~\cite{Chen:2002th}.
In this work, a similar strategy is extended to the penguin-dominated four-body decays
$B \rightarrow \phi(\rightarrow K\bar K) K^*(\rightarrow K\pi)$,
which  exhibit particularly alluring experimental and theoretical features.

In the SM, the decay $B \rightarrow \phi K^{*}$ is described by loop mediated $b\rightarrow d$
 or $b\rightarrow s$ transitions,
providing a sensitive test for NP.
The $B^0\rightarrow \phi K^{*0}$ decay was first observed by the CLEO Collaboration~\cite{CLEO:2001ium}.
Subsequently, branching ratio measurements and angular analyses have been reported by the $BABAR$ and Belle
Collaborations~\cite{BaBar:2003zor,BaBar:2004uwv,BaBar:2006ttd,BaBar:2008lan,Belle:2003ike,Belle:2005lvd,Belle:2013vat,Belle-II:2020rdz}.
The branching ratio as averaged by the Particle Data Group (PDG) is  $(1.00\pm 0.05)\times 10^{-5}$~\cite{pdg2020},
in good agreement with theoretical predictions~\cite{Beneke:2006hg,Cheng:2008gxa}.
The observed surprisingly large transverse polarization fractions, 
contrary to naive predictions based on helicity arguments (the so-called polarization puzzle~\cite{pdg2020}),
attracts much theoretical attention, with several explanations  proposed~\cite{Chen:2002pz,Li:2004mp,Grossman:2003qi,Das:2004hq,Hou:2004vj,Ladisa:2004bp,Colangelo:2004rd,
Bauer:2004tj,Kagan:2004uw,Chen:2005mka,Zou:2005gw,Yang:2004pm,Cheng:2004ru,Li:2004ti,Beneke:2005we,Datta:2007qb,Chang:2006dh,Bao:2008hd}.
Including both the $S$-wave $K^+\pi^-$ and $K^+K^-$ contributions,
the LHCb Collaboration measured the polarization amplitudes and $CP$ asymmetries
in $B^0\rightarrow \phi(\rightarrow K^+ K^-) K^{*0}(\rightarrow K^+\pi^-)$ decay~\cite{LHCb:2014xzf}.
The angular analysis was used to determine TPAs for the first time.
The measured true asymmetries show  no significant deviations from zero,
while several significant fake TPAs are consistent with the presence of final-state interactions.
As for the  $B_s$ counterpart,
the first observation of the decay $B_s^0\rightarrow \phi \bar{K}^{*0}$,
with $\phi\rightarrow K^+K^-$ and $\bar{K}^{*0}\rightarrow K^-\pi^+$,
was reported by the LHCb experiment~\cite{LHCb:2013nlu},
meanwhile, the determination of its branching ratio and polarizations were presented,
and the $S$-wave contribution  was estimated to be in the teens.
The measured value of the branching ratio is significantly larger than the theoretical predictions~\cite{Beneke:2006hg,Ali:2007ff,Cheng:2009mu,Zou:2015iwa,Wang:2017rmh,Wang:2017hxe,Yan:2018fif}.

As discussed above, we focus on the four-body decays $B\rightarrow \phi(\rightarrow K\bar K) K^*(\rightarrow K\pi)$,
where  the invariant mass of the $K\bar K$ ($K\pi$) pair is restricted to be within $\pm 15$ MeV ($\pm150$ MeV)
of the known mass of the $\phi(K^*(892))$ meson for comparison with the LHCb data.
 Except for the dominant vector resonances, two important scalar backgrounds,
 such as $f_0(980)\rightarrow K\bar K$ and $K^*_0(1430)\rightarrow K\pi$,
are also taken into account.
The contributions from higher spin resonances are expect to be small in the concerned mass regions
and are thus neglected in the following analysis.
Six different quasi-two-body decay channels are considered,
corresponding  to various different possible combinations of $K\bar{K}$ and $K\pi$ pairs with spin 0 and 1.
The $S$ and $P$-wave contributions are parametrized into the corresponding timelike form factors involved in the two-meson DAs,
which are well established in the three-body $B$ decays~\cite{Rui:2019yxx,Li:2021cnd}.
With these universal nonperturbative quantities,
we can make quantitative predictions on the various observables
including  the branching ratios, $S$-wave fractions, polarization fractions, direct $CP$ violations, and the TPAs
in $B\rightarrow \phi(\rightarrow K\bar K) K^*(\rightarrow K\pi)$ decays.

The rest of the paper is organized as follows.
Section~\ref{sec:framework} includes a general description of
the angular distribution,  kinematics, and the two-meson distribution amplitudes of the considered four-body decays.
We then apply the PQCD formalism in Sec.~\ref{sec:discussion} to the $B\rightarrow \phi(\rightarrow K\bar K) K^*(\rightarrow K\pi)$decays.
The numerical results are discussed and compared with those of other works in the literature.
Section~\ref{sec:sum} contains our conclusions.
The relevant factorization formulas  are collected in the Appendix~\ref{sec:ampulitude}.

\section{kinematics and two-meson distribution amplitudes}\label{sec:framework}
\subsection{Angular distribution and the helicity amplitudes}

\begin{figure}[!htbh]
\begin{center}
\vspace{0.01cm} \centerline{\epsfxsize=15cm \epsffile{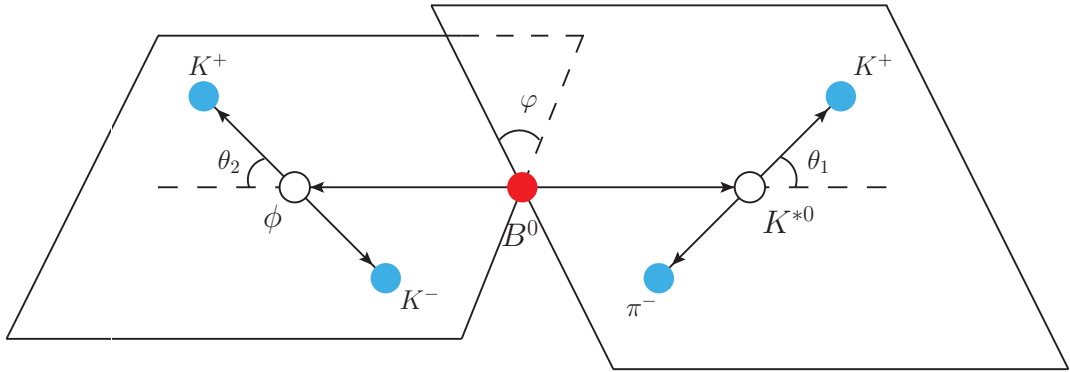}}
\setlength{\abovecaptionskip}{-13.5cm}
\caption{The definition of the decay angles $\theta_1$, $\theta_2$, and  $\varphi$ for the
decay $B^0\rightarrow \phi K^{*0}$ with $\phi\rightarrow K^+K^-$ and $K^{*0}\rightarrow K^+\pi^-$.
The angles are described in the text.}
\label{fig:Angle}
\end{center}
\end{figure}
The angular distribution in the $B^0\rightarrow \phi K^{*0}$ decay with $\phi\rightarrow K^+K^-$ and $K^{*0}\rightarrow K^+\pi^-$
is described by three angles $\theta_1$, $\theta_2$, and $\phi$ in the helicity basis, which are depicted in Fig.~\ref{fig:Angle}.
$\theta_1$  is the polar angle of the $K^+$ in the rest frame of the $K^*$ with respect to the helicity axis.
Similarly,  $\theta_2$ is the polar angle of the $K^+$ in the $\phi$ rest frame with respect to the helicity axis of the $\phi$.
The azimuth angle $\varphi$ is the relative angle between the $K^+ K^-$ and $K^+\pi^-$ decay planes in the $B$ rest frame.

Angular momentum conservation, for the vector-vector modes,
allows three possible polarization configurations of the $K \bar K$ and $K\pi$ pairs, such as longitudinal, parallel, or perpendicular.
The corresponding amplitudes are  denoted by $A_0$, $A_{\|}$, and $A_{\perp}$ respectively,
following the definitions given in Ref.~\cite{zjhep}.
Some scalar  resonances, such as $f_0(980)$ and $K_0^*(1430)$, are expected to contribute and thus are included
in the selected region of $K\bar K$ or $K\pi$ invariant masses.
As one of the meson pair is produced in a spin-0 ($S$-wave) state,
the resultant two  single $S$-wave amplitudes are  denoted as $A_{SV}$ and $A_{VS}$,
which are  physically different.
The double $S$-wave amplitude $A_{SS}$ is associated with the final state, where both meson pairs are produced in the $S$ wave.
All of the above considered decay modes,
together with  the corresponding amplitudes, are shown in Table \ref{tab:sp}.
\begin{table}
\caption{Quasi-two-body decay channels and the corresponding helicity amplitudes contributing
to the $(K\bar K)(K\pi)$ final state.
The subscript $S/P$ denotes an $S$- or $P$-wave configuration of the meson pair.}
\label{tab:sp}
\begin{tabular}[t]{lcc}
\hline\hline
Quasi-Two-Body Modes               & Resonance Types & Allowed Helicity Amplitudes \\ \hline
$B\rightarrow (K\bar K)_P(K\pi)_P$ & vector-vector   & $A_{0,\parallel,\perp}$     \\ 
$B\rightarrow (K\bar K)_S(K\pi)_P$ & scalar-vector   & $A_{SV}$                    \\ 
$B\rightarrow (K\bar K)_P(K\pi)_S$ & vector-scalar   & $A_{VS}$                    \\ 
$B\rightarrow (K\bar K)_S(K\pi)_S$ & scalar-scalar   & $A_{SS}$                    \\ 
\hline\hline
\end{tabular}
\end{table}

As discussed in Ref.~\cite{zjhep}, in the PQCD approach, these helicity amplitudes are expressed as the
convolution of the hard kernels with
the two-meson DAs, which absorb the nonperturbative dynamics involved in the meson pairs.
After regularizing the end-point singularities
and smearing the double logarithmic divergence, the typical factorization formula in coordinate space reads as
\begin{eqnarray}\label{eq:ampli1}
A\propto \int d x_B dx_1dx_2 b_B db_B b_1db_1b_2db_2 {\rm Tr}
[C(t)\Phi_B(x_B,b_B)\Phi_{K\pi}(x_1)\Phi_{K\bar K}(x_2)H(x_i,b_i,t)S_t(x_i)e^{-S(t)}],
\end{eqnarray}
where $x_i$ and $b_i$ with $i=B,1,2$ are the parton momentum fractions and  the conjugate space coordinate of the transverse momentum, respectively.
The threshold function $S_t(x_i)$ and the Sudakov exponents  $S(t)$ are given in the Appendix of Ref.\cite{zjhep}.
$t$ is the largest energy scale in hard function $H$.
$C(t)$ is the short distance Wilson coefficients at the hard scale $t$.
``Tr" denotes the trace over all Dirac structure and color indices.
The explicit analytic formulas for the considered helicity amplitudes are presented in the Appendix.

\subsection{The kinematics of four-body decay}

\begin{figure}[!htbh]
\begin{center}
\vspace{0.01cm} \centerline{\epsfxsize=7cm \epsffile{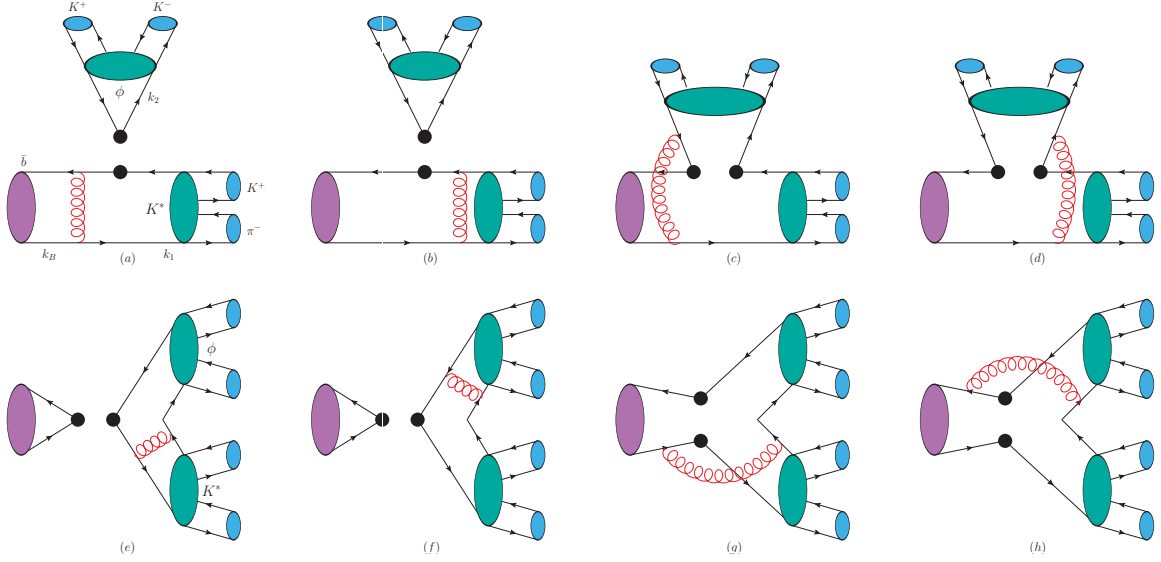}}
\setlength{\abovecaptionskip}{0cm}
\caption{Leading-order diagrams for the $B \rightarrow \phi(\rightarrow K\bar K) K^*(\rightarrow K\pi)$ decays, where the symbol
$\bullet$ denotes a weak vertex. }
\label{fig:Feynman}
\end{center}
\end{figure}

Taking the full cascade decay $B \rightarrow \phi(\rightarrow K\bar K) K^*(\rightarrow K\pi)$  as an example,
the kinematics can be described in terms of five independent variables:
three helicity angles ($\theta_1,\theta_2,\varphi$) and two invariant masses ($m_{KK}, m_{K\pi}$).
We first consider the subprocess $B \rightarrow \phi K^*$,
where $\phi$ and $K^*$ go subsequently into $K\bar K$ and $K\pi$ pairs, respectively.
Following the definition given in Ref.~\cite{zjhep},
the external momenta of the decay chain will be denoted as $p$, $q$ for the two meson pairs, and $p_{1,2}^{(')}$
for the four final-state mesons, with the specific charge assignment according to
\begin{eqnarray}
B(p_B)\rightarrow \phi(q)K^*(p)\rightarrow K(p_2)\bar K(p_2')K(p_1)\pi(p_1'),
\end{eqnarray}
where $p_B=p+q$, $p=p_1+p_1'$, and $q=p_2+p_2'$,
which obey the momentum conservation.
For simplicity, we shall work in the rest frame of the $B$ meson in the light-cone coordinates
such that $ p_B=\frac{M}{\sqrt{2}}(1,1,\textbf{0}_{T})$ with the $B$ meson mass $M$.
The momenta of $\phi$ and $K^*$ can be written as
\begin{eqnarray}\label{eq:pq}
q=\frac{M}{\sqrt{2}}(f^-,f^+,\textbf{0}_{T}),\quad p=\frac{M}{\sqrt{2}}(g^+,g^-,\textbf{0}_{T}),
\end{eqnarray}
respectively.
The factors $g^\pm$ and $f^\pm$ are related to the invariant masses of the meson pairs via $p^2=\omega_1^2$ and $q^2=\omega_2^2$,
which can be written as
\begin{eqnarray}
g^\pm&=&\frac{1}{2}[1+\eta_1-\eta_2\pm\sqrt{(1+\eta_1-\eta_2)^2-4\eta_1}], \nonumber\\
f^\pm&=&\frac{1}{2}[1-\eta_1+\eta_2\pm\sqrt{(1+\eta_1-\eta_2)^2-4\eta_1}],
\end{eqnarray}
with the mass ratios $\eta_{1,2}=\omega_{1,2}^2/M^2$.
For the  meson pairs in the $P$-wave configurations,
the corresponding longitudinal polarization vectors are defined as
 \begin{eqnarray}\label{eq:pq1}
\epsilon_{q}=\frac{1}{\sqrt{2\eta_2}}(-f^-,f^+,\textbf{0}_{T}),\quad
\epsilon_{p}=\frac{1}{\sqrt{2\eta_1}}(g^+,-g^-,\textbf{0}_{T}),
\end{eqnarray}
which satisfy the normalization $\epsilon_{q}^2=\epsilon_{p}^2=-1$  and the orthogonality
$\epsilon_{q}\cdot q=\epsilon_{p}\cdot p=0$.

Introducing the  meson momentum fraction $\zeta$ for each meson pair,
the individual momenta of the four final states can be expressed as
\begin{eqnarray}\label{eq:p1p2}
p_1&=&\left(\frac{M}{\sqrt{2}}(\zeta_1+\frac{r_1-r_1'}{2\eta_1})g^+,\frac{M}{\sqrt{2}}(1-\zeta_1+\frac{r_1-r_1'}{2\eta_1})g^-,\textbf{p}_{T}\right),\nonumber\\
p_1'&=&\left(\frac{M}{\sqrt{2}}(1-\zeta_1-\frac{r_1-r_1'}{2\eta_1})g^+,\frac{M}{\sqrt{2}}(\zeta_1-\frac{r_1-r_1'}{2\eta_1})g^-,-\textbf{p}_{T}\right),\nonumber\\
p_2&=&\left(\frac{M}{\sqrt{2}}(1-\zeta_2+\frac{r_2-r_2'}{2\eta_2})f^-,\frac{M}{\sqrt{2}}(\zeta_2+\frac{r_2-r_2'}{2\eta_2})f^+,\textbf{q}_{T}\right),\nonumber\\
p_2'&=&\left(\frac{M}{\sqrt{2}}(\zeta_2-\frac{r_2-r_2'}{2\eta_2})f^-,\frac{M}{\sqrt{2}}(1-\zeta_2-\frac{r_2-r_2'}{2\eta_2})f^+,-\textbf{q}_{T}\right),
\end{eqnarray}
with the mass ratios $r_i^{(\prime)}=m_i^{(\prime)2}/M^2(i=1,2)$, where $m_i^{(\prime)}$ is the mass of the meson $P_i^{(\prime)}$.
The transverse momenta $\textbf{p}_{T}$ and $\textbf{q}_{T}$
can be derived from the on-shell condition $p_i^{(\prime)2}=m_i^{(\prime)2}$ for each final-state meson, which yields
\begin{eqnarray}\label{eq:trans}
|\textbf{p}_{T}|^2=\omega_1^2[\zeta_1(1-\zeta_1)+\alpha_1],\quad |\textbf{q}_{T}|^2=\omega_2^2[\zeta_2(1-\zeta_2)+\alpha_2],
\end{eqnarray}
with the factors
\begin{eqnarray}
\alpha_{i}=\frac{(r_{i}-r_{i}')^2}{4\eta^2_{i}}-\frac{r_{i}+r_{i}'}{2\eta_{i}}.
\end{eqnarray}
Comparing Eqs.~(\ref{eq:pq}) and~(\ref{eq:p1p2}), we find the  meson momentum fractions modified by the meson masses,
\begin{eqnarray}
\frac{p_{1}^+}{p^+}=\zeta_{1}+\frac{r_{1}-r_{1}'}{2\eta_{1}},\quad
\frac{p_{2}^-}{q^-}=\zeta_{2}+\frac{r_{2}-r_{2}'}{2\eta_{2}}.
\end{eqnarray}
It is easy to derive the relation between $\zeta$ and  the polar angles $\theta$ in Fig.~\ref{fig:Angle} in the meson-pair rest frame:
 \begin{eqnarray}\label{eq:cos}
2\zeta_{i}-1&=& \sqrt{1+4\alpha_{i}}\text{cos} \theta_{i},
 \end{eqnarray}
 with the bound
  \begin{eqnarray}
 \zeta_{i}\in \Big[\frac{1-\sqrt{1+4\alpha_{i}}}{2},\frac{1+\sqrt{1+4\alpha_{i}}}{2}\Big].
 \end{eqnarray}
Note that Eq.~(\ref{eq:cos})
reduces to the conventional form in Refs.~\cite{npb905373,prd96051901} in the limit of massless.

The Feynman diagrams for the hard kernels associated with the considered four-body $B$ meson decays
are displayed in Fig.~\ref{fig:Feynman}, each of which contains a single hard gluon exchange
at leading order in the PQCD approach.
The first row represents the emission type,
while the second row represents the annihilation one.
Each type is further classified as factorizable,
in which a gluon attaches to quarks in the same meson,
and nonfactorizable, in which a gluon attaches to quarks in distinct mesons.
For the evaluation of the hard kernels,
we define three valence quark momenta labeled by $k_B$, $k_1$, and $k_2$ in Fig.~\ref{fig:Feynman}(a) as
 \begin{eqnarray}\label{eq:kt}
k_B&=&(0,x_Bp_B^-,\textbf{k}_{BT}),\quad k_1=(x_1p^+,0,\textbf{k}_{1T}),\quad k_2=(0,x_2q^-,\textbf{k}_{2T}),
\end{eqnarray}
with the parton momentum fractions $x_i$, and the parton transverse momenta $k_{iT}$.
Since $k_1$ and $k_2$ move with the corresponding  meson pair in the plus and minus direction, respectively,
the minus (plus) component of $k_1(k_2)$ can be neglected due to its small size.
We also drop $k_B^+$ because it does not appear in the hard kernels for dominant factorizable contributions.

\subsection{Two-meson distribution amplitudes}
The light-cone matrix elements for $S$-wave $K\bar K$ and $K\pi$ can be decomposed, up to twist 3, into~\cite{Chen:2002th,Wang:2015uea}
\begin{eqnarray}
\Phi_{(K\bar K)_S}(x_2,\omega_2)&=&\frac{1}{\sqrt{2N_c}}[\slashed{q}\phi_{(K\bar K)_S}^0(x_2,\omega_2)
+\omega_2\phi_{(K\bar K)_S}^s(x_2,\omega_2)+\omega_2(\slashed{v}\slashed{n}-1)\phi_{(K\bar K)_S}^t(x_2,\omega_2)],\nonumber\\
\Phi_{(K\pi)_S}(x_1,\omega_1)&=&\frac{1}{\sqrt{2N_c}}[\slashed{p}\phi_{(K\pi)_S}^0(x_1,\omega_1)
+\omega_1\phi_{(K\pi)_S}^s(x_1,\omega_1)+\omega_1(\slashed{n}\slashed{v}-1)\phi_{(K\pi)_S}^t(x_1,\omega_1)],
\end{eqnarray}
where $n=(1,0,\textbf{0}_{T})$ and $v=(0,1,\textbf{0}_{T})$ are two dimensionless vectors. 
The parametrization of various twists DAs  take the forms
\begin{eqnarray}\label{eq:phi0st}
\phi^0_{(PP')_S}(x,\omega)&=&
\left\{
\begin{aligned}
&\frac{9 F_{(PP')_S}(\omega)}{\sqrt{2N_c}}a_{PP'}x(1-x)(1-2x),   &PP'=K \bar K, \\
&\frac{3 F_{(PP')_S}(\omega)}{\sqrt{2N_c}}x(1-x) \left[\frac{1}{\mu_S}+B_13(1-2x)
+B_3 \frac{5}{2}(1-2x)(7(1-2x)^2-3)\right],  &PP'=K\pi, \\
\end{aligned}\right. \nonumber\\
\phi^s_{(PP')_S}(x,\omega)&=&\frac{ F_{(PP')_S}(\omega)}{2\sqrt{2N_c}},\nonumber\\
\phi^t_{(PP')_S}(x,\omega)&=&\frac{ F_{(PP')_S}(\omega)}{2\sqrt{2N_c}}(1-2x),
\end{eqnarray}
with the ratio $\mu_S=\omega_1/(m_{s}-m_{q})$, where $m_s(1\;{\rm GeV})=119$ MeV~\cite{prd73014017,prd77014034}
is the running strange quark mass and the light quark masses $m_q$, $q=u,d$, are set to zero.
The values of the Gegenbauer moments for the twist-2 DAs are taken as~\cite{Rui:2019yxx,Jia:2021uhi,prd73014017,prd77014034}
\begin{eqnarray}\label{eq:swavegen}
a_{K\bar K}=0.80\pm0.16, \quad B_1=-0.57\pm 0.13,\quad B_3=-0.42\pm 0.22.
\end{eqnarray}
Because the available data are not yet precise enough to extract more Gegenbauer moments,
the two twist-3 DAs are chosen as asymptotic forms.

For the scalar form factor $F_{(K\pi)_S}$, we follow the LASS line shape~\cite{npb296493},
which includes an effective-range nonresonant component with the $K_0^*(1430)$ resonance Breit-Wigner tail.
The explicit expression is given by
\begin{eqnarray}\label{eq:sform}
F_{(K\pi)_S}(\omega)&=&\frac{\omega}{k(\omega)}\cdot\frac{1}{\cot \delta_B-i}+
e^{2i \delta_B}\frac{m_0^2\Gamma_0/k(m_0)}{m_0^2-\omega^2-im_0^2
\frac{\Gamma_0}{\omega}\frac{k(\omega)}{k(m_0)}}, \nonumber\\
\cot\delta_B&=&\frac{1}{ak(\omega)}+\frac{1}{2}bk(\omega),
\end{eqnarray}
where  $m_0$ ($\Gamma_0$) is the mass (width) of $K^*_0(1430)$.
The kaon three-momentum $k(\omega)$ is written, in the $K\pi$ center-of-mass frame, as
\begin{eqnarray}\label{eq:kq}
k(\omega)=\frac{\sqrt{[\omega^2-(m_K+m_{\pi})^2][\omega^2-(m_K-m_{\pi})^2]}}{2\omega},
\end{eqnarray}
with $m_{K(\pi)}$ the known kaon (pion) mass,
and $k(m_0)$ being the same quantity evaluated at the nominal resonance mass $m_0$.
We use the following values for the resonance mass, width, scattering length, and effective-range parameters:
$m_0=1450\pm80$ MeV, $\Gamma_0=400\pm230$ MeV, $a=3.2\pm1.8$ GeV$^{-1}$, and $b=0.9\pm1.1$ GeV$^{-1}$~\cite{prd92012012,zjhep}.

For the scalar form factor of the $K \bar K$ system, the main resonance is $f_0(980)$ in the concerned mass window,
since its mass is very close to the $K \bar K$ threshold, which can strongly influence the resonance shape.
We follow Ref.~\cite{Rui:2019yxx} to take the the widely accepted prescription proposed by Flatt\'{e}~\cite{Flatte:1976xv}
\begin{eqnarray}
F_{(K\bar K)_S}(\omega)&=&\frac{m^2_{f_0(980)}}{m^2_{f_0(980)}-\omega^2-i m_{f_0(980)}(g_{\pi\pi}\rho_{\pi\pi}+g_{KK}\rho_{KK}F_{KK}^2)}
\end{eqnarray}
with $m_{f_{0}(980)}=939 {\rm MeV}/c^2, g_{\pi\pi}=199 {\rm MeV}/c^2, g_{KK}=3 g_{\pi\pi}$~\cite{LHCb:2014xzf}.
The exponential term $F_{KK}=e^{-\alpha q_K^2}$
is introduced above the $K\bar{K}$ threshold to reduce the $\rho_{KK}$ factor as $\omega$ increases,
where $q_k$ is the momentum of the kaon in the $K\bar{K}$ rest frame and $\alpha=2.0\pm 0.25$ GeV$^{-2}$~\cite{Aaij:2014emv,Bugg:2008ig}.

In Ref.~\cite{zjhep} we have updated the $P$-wave DAs including
both longitudinal and transverse polarizations for the $K\pi$ pair,
whose  explicit expressions read
\begin{eqnarray}
\Phi_{(K\pi)_P}^L(x_1,\zeta_1,\omega_1)&=&\frac{1}{\sqrt{2N_c}}[\omega_1\slashed{\epsilon}_p\phi_{(K\pi)_P}^0(x_1,\omega_1)
+\omega_1\phi_{(K\pi)_P}^s(x_1,\omega_1)
+\frac{\slashed{p}_1\slashed{p}'_{1}-\slashed{p}'_{1}\slashed{p}_{1}}{\omega_1(2\zeta_1-1)}\phi_{(K\pi)_P}^t(x_1,\omega_1)](2\zeta_1-1), \nonumber\\
\Phi_{(K\pi)_P}^T(x_1,\zeta_1,\omega_1)&=&\frac{1}{\sqrt{2N_c}}[\gamma_5\slashed{\epsilon}_T\slashed{p}\phi_{(K\pi)_P}^T(x_1,\omega_1)
+\omega_1\gamma_5\slashed{\epsilon}_T\phi_{(K\pi)_P}^a(x_1,\omega_1)+i\omega_1\frac{\epsilon^{\mu\nu\rho\sigma}\gamma_\mu
\epsilon_{T\nu}p_\rho n_{-\sigma}}{p\cdot n_-}\phi_{(K\pi)_P}^v(x_1,\omega_1)]\nonumber\\ &&\sqrt{\zeta_1(1-\zeta_1)+\alpha_1},
\end{eqnarray}
respectively.
Naively, the $P$-wave $K\bar K$ ones can be obtained with the following replacement:
\begin{eqnarray}
x_1\rightarrow x_2, \omega_1\rightarrow \omega_2, \zeta_1\rightarrow \zeta_2, \alpha_1\rightarrow \alpha_2,
p\rightarrow q, \epsilon_{p}\rightarrow \epsilon_{q},
p_1^{(')}\rightarrow p_2^{(')}.
\end{eqnarray}

The various twist  DAs for the $P$-wave $K\bar K$ and $K\pi$ systems are parametrized as~\cite{Li:2021cnd}
\begin{eqnarray}
\phi_{K\bar{K}}^0(x_2,\omega_2)&=&\frac{3}{\sqrt{2N_c}}F^{\parallel}_{K\bar{K}}(\omega_2)x_2(1-x_2)
[1+{a_2^0}_{\phi}C_2^{3/2}(2x_2-1)], \nonumber\\
\phi_{K\bar{K}}^s(x_2,\omega_2)&=&\frac{3}{2\sqrt{2N_c}}F^{\perp}_{K\bar{K}}(\omega_2)(1-2x_2), \nonumber\\
\phi_{K\bar{K}}^t(x_2,\omega_2)&=&\frac{3}{2\sqrt{2N_c}}F^{\perp}_{K\bar{K}}(\omega_2)(1-2x_2)^2,\nonumber\\
\phi_{K\bar{K}}^T(x_2,\omega_2)&=&\frac{3}{\sqrt{2N_c}}F^{\perp}_{K\bar{K}}(\omega_2)x_2(1-x_2)
[1+{a_2^0}_{\phi}C_2^{3/2}(2x_2-1)], \nonumber\\
\phi_{K\bar{K}}^a(x_2,\omega_2)&=&\frac{3}{4\sqrt{2N_c}}F^{\parallel}_{K\bar{K}}(\omega_2)(1-2x_2), \nonumber\\
\phi_{K\bar{K}}^v(x_2,\omega_2)&=&\frac{3}{8\sqrt{2N_c}}F^{\parallel}_{K\bar{K}}(\omega_2)[1+(1-2x_2)^2], \nonumber\\
\phi_{K\pi}^0(x_1,\omega_1)&=&\frac{3}{\sqrt{2N_c}}F^{\parallel}_{K\pi}(\omega_1)x_1(1-x_1)
[1+{a_1^0}_{K^*}C_1^{3/2}(2x_1-1)+{a_2^0}_{K^*}C_2^{3/2}(2x_1-1)], \nonumber\\ 
\phi_{K\pi}^s(x_1,\omega_1)&=&\frac{3}{2\sqrt{2N_c}}F^{\perp}_{K\pi}(\omega_1)(1-2x_1), \nonumber\\
\phi_{K\pi}^t(x_1,\omega_1)&=&\frac{3}{2\sqrt{2N_c}}F^{\perp}_{K\pi}(\omega_1)(1-2x_1)^2, \nonumber\\
\phi_{K\pi}^T(x_1,\omega_1)&=&\frac{3}{\sqrt{2N_c}}F^{\perp}_{K\pi}(\omega_1)x_1(1-x_1)
[1+{a_1^0}_{K^*}C_1^{3/2}(2x_1-1)+{a_2^0}_{K^*}C_2^{3/2}(2x_1-1)], \nonumber\\ 
\phi_{K\pi}^a(x_1,\omega_1)&=&\frac{3}{4\sqrt{2N_c}}F^{\parallel}_{K\pi}(\omega_1)(1-2x_1), \nonumber\\
\phi_{K\pi}^v(x_1,\omega_1)&=&\frac{3}{8\sqrt{2N_c}}F^{\parallel}_{K\pi}(\omega_1)[1+(1-2x_1)^2],
\end{eqnarray}
with the Gegenbauer polynomials
\begin{eqnarray}
C_1^{3/2}(t)=3t, \quad C_2^{3/2}(t)=\frac{3}{2}(5t^2-1). 
\end{eqnarray}
The values of the Gegenbauer moments associated with longitudinal polarization are adopted as
\begin{eqnarray}\label{eq:pwavegen}
{a_1^0}_{K^*}=0.31\pm0.16, \quad {a_2^0}_{K^*}=1.19\pm0.10, \quad {a_2^0}_{\phi}=-0.31\pm0.19, 
\end{eqnarray}
which are determined from a global analysis in the PQCD approach~\cite{Li:2021cnd}.
We do not distinguish the Gegenbauer moments for the longitudinal and transverse
polarizations in our numerical calculations due to a lack of rigorous theoretical
 and experimental information on the transverse polarizations.

Since the $K\pi$ spectrum is dominated by the vector $K^*(892)$ resonance in the selected invariant mass range,
the $P$-wave timelike form factor $F_{K\pi}^{\parallel}$ is parametrized as the relativistic Breit-Wigner model~\cite{cpc44073102,prd101016015}
\begin{eqnarray}
F_{K\pi}^{\parallel}(\omega)=\frac{m_{K^*}^2}{m_{K^*}^2-\omega^2-im_{K^*}\Gamma(\omega)},
\end{eqnarray}
where  $m_{K^*}=895.81$ MeV is the $K^*(892)$ mass.
The mass-dependent width is given by
\begin{eqnarray}
\Gamma(\omega)=\Gamma_{K^*}\frac{k^3(\omega)}{k^3(m_{K^*})}\frac{m_{K^*}}{\omega}
\frac{1+r^2k^2(m_{K^*})}{1+r^2k^2(\omega)},
\end{eqnarray}
where $\Gamma_{K^*}=47.4$ MeV is the natural width of the $K^*(892)$ meson and $r=3.4$ GeV$^{-1}$ is the interaction radius~\cite{LHCb:2014xzf}.
The $P$-wave $K\bar K$ one,
denoted $F_{K\bar K}^{\parallel}$, is modeled in a similar way using the values  $m_{\phi}=1019.455$ MeV and $\Gamma_{\phi}=4.26$ MeV~\cite{LHCb:2014xzf}.
For another form factor $F^{\perp}$, we assume the approximate relation
$F^{\perp}/ F^{\parallel}\sim f^{T}_{V}/f_{V}$  with $f^{(T)}_{V}$ being the tensor (vector) decay constant of the vector resonance.

\section{numerical results}\label{sec:discussion}
In this section we discuss in detail some physical observables, such as branching ratios, $S$-wave
fractions, polarization fractions, direct $CP$ asymmetries, and TPAs, for the concerned decays.
The related input parameters for the numerical calculations are collected in Table~\ref{tab:constant}.
The decay constants used the values from Refs.~\cite{zjhep,Ali:2007ff,Li:2021cnd}, while the meson masses, Wolfenstein parameters,
and the lifetimes are taken from the PDG review~\cite{pdg2020}.
We neglect uncertainties on the constants since they are negligible with respect to other sources of uncertainty.
The parameters relevant to the $K\bar K$ and $K\pi$ DAs have been specified in  the previous section.

Another key input in the PQCD calculations is the $B$ meson distribution amplitude.
We adopt the conventional form from Refs.~\cite{Li:2003yj,Kurimoto:2001zj} of the leading Lorentz structure
\begin{eqnarray}
\phi_{B}(x,b)=N_B x^2(1-x)^2\exp\big[-\frac{x^2M^2}{2\omega^2_b}-\frac{\omega^2_bb^2}{2}\big],
\end{eqnarray}
with the shape parameter $\omega_b=0.40$ GeV for $B_{u,d}$ mesons and
$\omega_b=0.48$ GeV for a $B_s$ meson~\cite{Hua:2020usv}. The normalization constant
$N_B$ is related to the decay constant $f_B$ via the normalization
\begin{eqnarray}
\int_0^1\phi_{B}(x,b=0)d x=\frac{f_{B}}{2\sqrt{2N_c}}.
\end{eqnarray}
For more alternative models of the $B$ meson DA and the  subleading contributions,
one can refer to Refs.~\cite{prd102011502,prd70074030,Li:2012md,Li:2014xda,Li:2012nk}.

\begin{table}
\caption{The decay constants are taken from Refs.~\cite{zjhep,Ali:2007ff,Li:2021cnd}.
Other parameters are from PDG 2020~\cite{pdg2020}. }
\label{tab:constant}
\begin{tabular*}{18cm}{@{\extracolsep{\fill}}llccc}
  \hline\hline
\text{Mass(\text{GeV})}
& $M_{B_s}=5.37$  & $M_B=5.28$  &$m_{K}=0.494$ & $m_{\pi}=0.14$ \\[1ex]
\end{tabular*}
\begin{tabular*}{18cm}{@{\extracolsep{\fill}}llccc}
\text{Wolfenstein parameters}
& $\lambda=0.22650$  & $A=0.790$  &$\bar{\rho}=0.141$ & $\bar{\eta}=0.357$ \\[1ex]
\end{tabular*}
\begin{tabular*}{18cm}{@{\extracolsep{\fill}}lcccccc}
\text{Decay constants (GeV)}
& $f_{B_s}=0.24$ & $f_{B}=0.21$  &$f_{\phi(1020)}=0.215$ &$f_{\phi(1020)}^T=0.186$ &$f_{K^*}=0.217$ &$f^T_{K^*}=0.185$ \\[1ex]
\end{tabular*}
\begin{tabular*}{18cm}{@{\extracolsep{\fill}}lccc}
\text{Lifetime (ps)}
& $\tau_{B_s}=1.51$ & $\tau_{B^0}=1.52$ & $\tau_{B^+}=1.638$ \\[1ex]
\hline\hline
\end{tabular*}
\end{table}
\subsection{$CP$ averaged four-body branching ratios and $S$-wave fractions}\label{sec:br}
The phase space of a four-body decay relies on the five kinematic variables,
that is, three helicity angles shown in Fig.~\ref{fig:Angle} and two invariant masses.
In the $B$ meson rest frame, the fivefold differential decay rate can be written as
\begin{eqnarray}\label{eq:dB}
\frac{d^5\Gamma}{d \theta_1 d\theta_2 d\varphi d\omega_1 d\omega_2}=\frac{k(\omega_1)k(\omega_2)k(\omega_1,\omega_2)}{16(2\pi)^6M^2}|A|^2,
\end{eqnarray}
where $k(\omega_1,\omega_2)=\sqrt{[M^2-(\omega_1+\omega_2)^2][M^2-(\omega_1-\omega_2)^2]}/(2M)$ is the magnitude of the three-momentum of the meson pair in the $B$ meson rest frame.
By appropriate variable changes, Eq.~(\ref{eq:dB}) is equivalent to the one in Ref.~\cite{plb770348}.
Replacing the helicity angles $\theta_i$ by the meson momentum fractions $\zeta_i$
via Eq.~(\ref{eq:cos}) and integrating the decay rate with respect to all independent variables, we obtain
\begin{eqnarray}\label{eq:dB2}
\Gamma= \frac{1}{4(2\pi)^6M^2\sqrt{1+4\alpha_1}\sqrt{1+4\alpha_2}}\int
k(\omega_1)k(\omega_2)k(\omega_1,\omega_2)|A|^2d \zeta_1 d\zeta_2 d\varphi d\omega_1 d\omega_2,
\end{eqnarray}
where the selected invariant mass ranges for the $K\bar K$ and $K\pi$ pairs
are $m_{\phi}-0.015<\omega_2<m_{\phi}+0.015$ (GeV) and $m_{K^*}-0.15<\omega_1<m_{K^*}+0.15$ (GeV), respectively.
The total amplitude ($A$)  can be decomposed into six helicity components shown in Table~\ref{tab:sp} 
with different  $\zeta_i$ and $\varphi$ dependencies~\cite{zjhep}
\begin{eqnarray}
A&=&\frac{2\zeta_1-1}{\sqrt{1+4\alpha_1}}\frac{2\zeta_2-1}{\sqrt{1+4\alpha_2}}A_0
+2\sqrt{2}\sqrt{\frac{\zeta_1(1-\zeta_1)+\alpha_1}{1+4\alpha_1}}\sqrt{\frac{\zeta_2(1-\zeta_2)+\alpha_2}{1+4\alpha_2}}\cos(\varphi)A_\parallel \nonumber\\&&
+i2\sqrt{2}\sqrt{\frac{\zeta_1(1-\zeta_1)+\alpha_1}{1+4\alpha_1}}\sqrt{\frac{\zeta_2(1-\zeta_2)+\alpha_2}{1+4\alpha_2}}\sin(\varphi)A_\perp
+\frac{2\zeta_1-1}{\sqrt{1+4\alpha_1}}A_{VS} \nonumber\\&&
+\frac{2\zeta_2-1}{\sqrt{1+4\alpha_2}}A_{SV}+A_{SS}.
\end{eqnarray}
The branching ratio of each component is then
\begin{eqnarray}\label{eq:dBh}
\mathcal{B}_h=\frac{\tau_B}{4(2\pi)^6M^2}\frac{2\pi}{9}Y_h\int d\omega_1d\omega_2k(\omega_1)k(\omega_2)k(\omega_1,\omega_2)|A_h|^2,
\end{eqnarray}
with
\begin{eqnarray}
Y_h = \begin{cases}
1, & h=0, \parallel, \perp\\
3, & h=SV, VS\\
9, & h=SS.
\end{cases}
\end{eqnarray}
which come from the integrations over $\zeta_1,\zeta_2,\varphi$.
Combining Eq.~(\ref{eq:dBh}) with its counterpart of the corresponding  $CP$-conjugated process,
we can derive the $CP$-averaged branching ratio of each component and their sum
\begin{eqnarray}
\mathcal{B}_h^{avg}=\frac{1}{2}(\mathcal{\bar{B}}_h+\mathcal{B}_h), \quad \mathcal{B}_{\text{total}}=\sum_h \mathcal{B}_h,
\end{eqnarray}
with $h$ running over the six helicities as stated above.

The numerical results are summarized in Table \ref{tab:br},
where the first quoted uncertainty is due to the shape parameters $\omega_b$ in the $B_{(s)}$ meson DAs with $10\%$ variation,
the second uncertainty is caused by the variation of the Gegenbauer moments in the two-meson DAs shown
in Eqs.~(\ref{eq:swavegen}) and (\ref{eq:pwavegen}),
and the last one
comes from  the hard scale $t$ that varies from $0.75t$ to $1.25t$
and the QCD scale $\Lambda_{QCD}=0.25\pm0.05 $ GeV.
The three uncertainties are comparable, and their combined impacts could exceed $50\%$,
implying that the nonperturbative parameters in the DAs of the initial and final states must be more precisely restricted,
and the higher-order correction to four-body $B$ meson decays is critical.
The concerned three channels are all penguin-dominated decays.
The $B^0$ and $B^+$ modes involve the $b\rightarrow s$ transition and
therefore have relatively large  branching ratios of $\mathcal{O}(10^{-7}\sim 10^{-6})$,
while the $B_s$ channels, mediated by the $b\rightarrow d$ transition, are generally 1 or 2 orders of magnitudes smaller.

\begin{table}[!htbh]
\caption{PQCD predictions for the $CP$-averaged branching ratios of various components and their
sum in the $B_{(s)}\rightarrow (K\bar K)(K\pi)$ decays
within the $K\bar K(K\pi)$ invariant mass window of 15 (150) MeV around the $\phi(K^*(982))$ resonance.}
\label{tab:br}
\begin{tabular}[t]{lccc}
\hline\hline
Components              & $B^0\rightarrow(K^+K^-)(K^+\pi^-)$               & $B_s^0\rightarrow(K^+K^-)(K^-\pi^+)$             & $B^+\rightarrow(K^+K^-)(K^0\pi^+)$               \\ \hline
$\mathcal{B}_0$         & $1.8_{-0.6-0.4-0.4}^{+0.7+0.3+0.6}\times10^{-6}$ & $3.1_{-0.8-0.9-1.2}^{+1.2+1.0+1.6}\times10^{-8}$ & $1.8_{-0.5-0.3-0.4}^{+0.9+0.4+0.7}\times10^{-6}$ \\
$\mathcal{B}_\parallel$ & $3.1_{-0.4-0.4-0.8}^{+0.5+0.4+1.3}\times10^{-7}$ & $5.3_{-0.3-1.4-1.3}^{+0.6+2.0+2.5}\times10^{-9}$ & $3.4_{-0.4-0.4-0.8}^{+0.6+0.5+1.3}\times10^{-7}$ \\
$\mathcal{B}_\perp$     & $3.3_{-0.4-0.4-0.8}^{+0.6+0.5+1.3}\times10^{-7}$ & $5.2_{-0.3-1.6-1.5}^{+0.3+1.9+2.6}\times10^{-9}$ & $3.6_{-0.5-0.5-0.9}^{+0.6+0.4+1.3}\times10^{-7}$ \\
$\mathcal{B}_{SS}$      & $4.7_{-1.4-1.9-1.4}^{+2.0+2.4+1.8}\times10^{-8}$ & $1.3_{-0.5-0.5-0.4}^{+0.7+0.6+0.5}\times10^{-9}$ & $5.2_{-1.5-2.1-1.6}^{+2.1+2.6+2.0}\times10^{-8}$ \\
$\mathcal{B}_{VS}$      & $4.6_{-0.9-1.0-1.4}^{+1.2+1.0+1.7}\times10^{-7}$ & $5.0_{-1.8-1.1-2.1}^{+2.3+1.2+2.7}\times10^{-9}$ & $4.8_{-1.1-1.0-1.3}^{+1.5+1.0+1.8}\times10^{-7}$ \\
$\mathcal{B}_{SV}$      & $2.2_{-0.5-0.4-0.6}^{+0.6+0.4+0.8}\times10^{-7}$ & $1.6_{-0.4-0.6-0.4}^{+0.5+0.8+0.5}\times10^{-9}$ & $2.5_{-0.5-0.4-0.7}^{+0.7+0.5+0.9}\times10^{-7}$ \\
$\mathcal{B}_{total}$   & $3.2_{-0.9-0.5-0.8}^{+1.0+0.4+1.1}\times10^{-6}$ & $4.9_{-1.1-1.2-1.6}^{+1.6+1.4+2.4}\times10^{-8}$ & $3.3_{-0.8-0.5-0.8}^{+1.2+0.6+1.2}\times10^{-6}$ \\
\hline\hline
\end{tabular}
\end{table}
Although the $P$-wave contributions dominate in the selected mass regions, the $S$-wave contributions,
which are strongly  sensitive to the integrating ranges,  cannot be neglected.
In order to compare the relative size of the
$S$-wave contributions, one can define the $S$-wave fractions as
\begin{eqnarray}\label{eq:fss}
f_{\sigma}&=&\frac{\mathcal{B}_{\sigma}}{\mathcal{B}_{\text{total}}},
\end{eqnarray}
with $\sigma=\{SS,SV,VS\}$, and the total $S$-wave fraction as $f_{\text{S-wave}}=f_{SS}+f_{SV}+f_{VS}$.
Using the PQCD predictions as given in Table~\ref{tab:br},
it is straightforward to obtain the numerical results of $S$-wave fractions in each channel
as exhibited in Table~\ref{tab:Sf}.

The predicted two single $S$-wave fractions for the $B^0$ mode
are well consistent with the data from Ref.~\cite{LHCb:2014xzf}.
In Ref.~\cite{LHCb:2014xzf}, it is assumed that the double $S$-wave component is negligible.
In our calculations, as can be seen in Table~\ref{tab:Sf}, the double $S$-wave fractions of all the three modes
are estimated to be less than $3\%$ and can be safely ignored. Therefore the assumption in~\cite{LHCb:2014xzf}
is reasonable in the selected invariant mass ranges.
However, we will discuss later that the $S$-wave contributions will be enhanced rapidly with increasing invariant mass ranges.
The predicted total $S$-wave fraction for the $B_s$ mode is in agreement with the LHCb data $0.16\pm 0.02$~\cite{LHCb:2013nlu}
obtained from  the total $\phi K^{*0}$ purity  by combining the $K^+K^-$ and $K^+\pi^-$ contributions.
It is assumed in Ref.~\cite{LHCb:2013nlu} that the $S$-wave component is the same for $B^0\rightarrow \phi K^{*0}$ and $B^0_s\rightarrow \phi \bar{K}^{*0}$ decays,
so that the  larger sample of $B^0\rightarrow \phi K^{*0}$ decays can be used.
This assumption lead to a large systematic uncertainty of this measurement,
which is expected to scale with larger data samples in future.
From Table~\ref{tab:Sf},
the total $S$-wave contributions could be as large as $20\%$ of the total decay rate,
indicating they are numerically significant in the given mass regions.

\begin{table}[!htbh]
\caption{$S$-wave fractions in the $B_{(s)}\rightarrow (K\bar K)(K\pi)$ decays
within the $K\bar K(K\pi)$ invariant mass window of 15 (150) MeV around the $\phi(K^*(982))$ resonance.
The data are taken from Ref.~\cite{LHCb:2014xzf},
where the first uncertainty is statistical and the second is systematic.}
\label{tab:Sf}
\begin{tabular}[t]{lcccc}
\hline\hline
Modes                                & $f_{SS}$                                          & $f_{VS}$                                          & $f_{SV}$                                          & $f_{S-wave}$   \\ \hline
$B^0\rightarrow(K^+K^-)(K^+\pi^-)$   & $0.015_{-0.000-0.006-0.002}^{+0.001+0.007+0.002}$ & $0.144_{-0.007-0.020-0.015}^{+0.013+0.022+0.016}$ & $0.069_{-0.002-0.011-0.005}^{+0.006+0.014+0.002}$ & $0.228_{-0.009-0.032-0.022}^{+0.020+0.037+0.018}$ \\
LHCb~\cite{LHCb:2014xzf}             & $\cdots$                                          & $0.143\pm0.013\pm0.012$                           & $0.122\pm0.013\pm0.008$                           & $\cdots$       \\ 
$B_s^0\rightarrow(K^+K^-)(K^-\pi^+)$ & $0.027_{-0.004-0.011-0.002}^{+0.004+0.013+0.002}$ & $0.102_{-0.018-0.030-0.029}^{+0.010+0.032+0.022}$ & $0.033_{-0.001-0.014-0.007}^{+0.000+0.016+0.004}$ & $0.162_{-0.023-0.050-0.037}^{+0.014+0.057+0.027}$ \\ 
$B^+\rightarrow(K^+K^-)(K^0\pi^+)$   & $0.016_{-0.001-0.006-0.002}^{+0.000+0.008+0.001}$ & $0.146_{-0.006-0.021-0.009}^{+0.003+0.021+0.014}$ & $0.076_{-0.005-0.013-0.007}^{+0.003+0.014+0.001}$ & $0.238_{-0.012-0.035-0.018}^{+0.006+0.035+0.015}$ \\
\hline\hline
\end{tabular}
\end{table}

The $S$-wave channels were also measured by other collaborations with a different $K\pi$ mass range.
The  branching ratio for $B^0\rightarrow \phi (K\pi)^*_0$ mode
was measured to be $(4.3\pm 0.4(\text{stat})\pm 0.4(\text{syst}))\times 10^{-6}$ by the Belle experiment~\cite{Belle:2013vat},
which was consistent with the previous $BABAR$ measurement, $(4.3\pm 0.6(\text{stat})\pm 0.4(\text{syst}))\times 10^{-6}$~\cite{BaBar:2008lan}.
We note that the above two measurements
were performed in a broad $K\pi$ invariant mass range $0.7<m_{K\pi}<1.55$ GeV.
For comparison, we derive the single $S$-wave branching ratio with the same mass region,
\begin{eqnarray}
\mathcal{B}(B^0\rightarrow \phi(\rightarrow K^+K^-) (K\pi)^*_0(\rightarrow K^+\pi^-))=(1.2_{-0.1-0.2-0.3}^{+0.4+0.3+0.5})\times 10^{-6},
\end{eqnarray}
which is more than  twice  the number in Table~\ref{tab:br}.
After correcting for the secondary branching fraction
$\mathcal{B}(\phi\rightarrow K^+K^-)=0.5$ and $\mathcal{B}((K\pi)^*_0 \rightarrow K^+\pi^-)=2/3$,
one can obtain the two-body decay branching ratio  in the narrow-width limit,
\begin{eqnarray}
\mathcal{B}(B^0\rightarrow \phi(K\pi)^*_0)=(3.6_{-0.3-0.6-0.9}^{+1.2+0.9+1.5})\times 10^{-6},
\end{eqnarray}
which complies with the above two measurements and
the previous two-body PQCD value of $(3.7^{+0.8+0.1+3.7}_{-0.7-0.1-1.7})\times 10^{-6}$ for the  S1 scenario~\cite{Kim:2009dg}.

In Ref.~\cite{BaBar:2011ryf}, the $BABAR$ Collaboration
measured the  $B^0\rightarrow f_0(980)K^{*0}(892)$ and $B^0\rightarrow f_0(980)(K\pi)^*_0$  decays,
where $f_0(980)$ was reconstructed through $f_0(980)\rightarrow \pi\pi$ within the $\pi\pi$
invariant mass range $0.47<m_{\pi\pi}<1.07$ GeV.
For the $K\pi$ mass spectrum, the two channels were analyzed separately.
 The former was performed in the ``low mass region(LMR)'', $0.75<m_{K\pi}<1.0$ GeV,
 while the latter was performed in the ``high mass region(HMR)'', $1.0<m_{K\pi}<1.55$ GeV.
The quoted branching ratios yield~\cite{BaBar:2011ryf}
\begin{eqnarray}\label{eq:ssvv}
\mathcal{B}(B^0\rightarrow f_0(980)(K\pi)^*_0)\times \mathcal{B}(f_0(980)\rightarrow \pi\pi)\times
\mathcal{B}((K\pi)^*_0\rightarrow K\pi)&=&(3.1\pm0.8\pm0.7)\times 10^{-6},\nonumber\\
\mathcal{B}(B^0\rightarrow f_0(980)K^{*0}(892))\times \mathcal{B}(f_0(980)\rightarrow \pi\pi)&=&(5.7\pm0.6\pm0.4)\times 10^{-6},
\end{eqnarray}
where the uncertainties are statistical and  systematic, respectively.
The Belle Collaboration  presented a smaller value of
$\mathcal{B}(B^0\rightarrow f_0(980)K^{*0}(892))=(1.4^{+0.6+0.6}_{-0.5-0.4})\times 10^{-6}$,
obtained for $m_{K\pi}\in (0.75,1.2)$ GeV and $m_{\pi\pi}\in (0.55,1.2)$ GeV,
with $2.5\sigma$ significance~\cite{Belle:2009roe}.
For comparison we recalculate the four-body branching ratios for the scalar-scalar and scalar-vector modes
  with the same $K\pi$ mass region as the $BABAR$ experiment.
In the narrow-width limit,  we obtain the products
\begin{eqnarray}\label{eq:ssvv1}
\mathcal{B}(B^0\rightarrow f_0(980)(K\pi)^*_0))\times \mathcal{B}(f_0(980)\rightarrow K^+K^-)
\times\mathcal{B}((K\pi)^*_0\rightarrow K^+\pi^-)&=&(2.1_{-0.6-0.7-0.6}^{+0.7+0.9+0.9})\times 10^{-7},\nonumber\\
\mathcal{B}(B^0\rightarrow f_0(980)K^{*0}(892))\times \mathcal{B}(f_0(980)\rightarrow K^+K^-)
\times \mathcal{B}(K^{*0}(892)\rightarrow K^+\pi^-)&=&(4.3_{-0.9-0.7-1.2}^{+1.3+0.7+1.6})\times 10^{-7}.
\end{eqnarray}
It is worth emphasizing that the above results cannot be directly compared with the data in Eq.~(\ref{eq:ssvv})
due to the absence of reliable information about $\mathcal{B}(f_0(980)\rightarrow K^+K^-)$ and $\mathcal{B}(f_0(980)\rightarrow \pi^+\pi^-)$.
Assuming that the $f_0(980)$ resonance only couples to the $K\bar K$ and $\pi\pi$ channels
and using the BES measurement~\cite{BES:2005iaq}
\begin{eqnarray}
\frac{\Gamma(f_0(980)\rightarrow \pi\pi)}{\Gamma(f_0(980)\rightarrow \pi\pi)+\Gamma(f_0(980)\rightarrow K\bar K)}=0.75^{+0.11}_{-0.13},
\end{eqnarray}
 and  isospin relations $\Gamma(f_0(980)\rightarrow \pi^+\pi^-)/\Gamma(f_0(980)\rightarrow \pi\pi)=2/3$ and
 $\Gamma(f_0(980)\rightarrow K^+K^-)/\Gamma(f_0(980)\rightarrow K\bar K)=1/2$, we obtain the ratio
 \begin{eqnarray}\label{eq:kkpi}
\frac{\Gamma(f_0(980)\rightarrow \pi^+\pi^-)}{\Gamma(f_0(980)\rightarrow K^+K^-)}=4.0^{+0.6}_{-0.7}.
\end{eqnarray}
Plugging Eq.~(\ref{eq:kkpi}) into Eq.~(\ref{eq:ssvv1}), and using above isospin relations,
we estimate the branching ratio products in Eq.~(\ref{eq:ssvv}) to be
$(1.9^{+1.4}_{-1.0})\times 10^{-6}$ and $(3.9^{+2.0}_{-1.5})\times 10^{-6}$, respectively,
where the errors are added in quadrature.
It can be seen  that our prediction for the former is  consistent with the $BABAR$ value within uncertainties,
but with central values that are somewhat lower.
The number of the latter  is in between  $BABAR$ and Belle measurements within errors.
As pointed out in Refs.~\cite{Cheng:2020iwk,Cheng:2020mna},
 the narrow-width approximation should be corrected by including finite-width effects for the broad scalar intermediate states.
The results of Eq.~(\ref{eq:ssvv1}), extracted from the four-body branching ratios,
may suffer from a large uncertainty due to the finite-width effects of the scalar resonance.
Thus the above comparisons is just a rough estimate for a cross-checking.
In addition, we have shown that  the $S$-wave contributions
depend on the range of the $K\bar K$ and $K\pi$ invariant masses.
In this study, the $K\bar K$ invariant mass
is limited in a narrow window of $\pm 15$ MeV around the known $\phi$ mass,
and a broader integrated region would increase our results.
It is expected that future experiments can directly reconstruct
intermediate $f_0(980)$ resonance through  $f_0(980)\rightarrow K^+K^-$ in the $B^0\rightarrow (K^+K^-)(K^+\pi^-)$ decay.

\subsection{Two-body branching ratios and polarization fractions}\label{sec:brplo}
\begin{table}[!htbh]
\caption{$CP$-averaged branching ratios and polarization fractions for the two-body $B\rightarrow \phi K^*$ decays.
For comparison, we also list the results from
PQCD \cite{Chen:2002pz, Ali:2007ff, Zou:2015iwa},
QCDF \cite{Beneke:2006hg, Cheng:2008gxa, Cheng:2009mu, Chang:2017brr},
SCET \cite{Wang:2017rmh}, and FAT \cite{Wang:2017hxe}.
The world averages of experimental data are taken from PDG 2020~\cite{pdg2020}.}
\label{tab:TwobodyBr}
\begin{tabular}[t]{lccc}
\hline\hline
Modes                               & $\mathcal{B}(10^{-6})$           & $f_0(\%)$            & $f_\perp(\%)$         \\ \hline
$B^0\rightarrow\phi K^{*0}$         & $7.4_{-2.1-1.2-1.8}^{+2.5+1.1+2.6}$ & $74.1_{-5.8-3.0-1.2}^{+3.1+1.5+1.1}$ & $13.3_{-1.6-0.8-0.5}^{+3.0+1.6+0.6}$ \\
PQCD-I~\cite{Chen:2002pz}           & $14.86$                          & $75.0$               & $11.5$                \\
PQCD-II \cite{Zou:2015iwa}          & $9.8^{+4.9}_{-3.8}$              & $56.5^{+5.8}_{-5.9}$ & $21.3^{+2.8}_{-2.9}$  \\
QCDF-I \cite{Beneke:2006hg}         & $9.3^{+0.5+11.4}_{-0.5-6.5}$     & $44^{+0+59}_{-0-36}$ & $\cdots$              \\
QCDF-II \cite{Cheng:2008gxa}        & $9.5^{+1.3+11.9}_{-1.2-5.9}$     & $50^{+50}_{-42}$     & $25^{+21}_{-25}$      \\
SCET \cite{Wang:2017rmh}            & $9.14\pm3.14$                    & $51.0\pm16.4$        & $22.2\pm9.9$          \\
FAT \cite{Wang:2017hxe}             & $8.64\pm1.76\pm1.70\pm0.90$      & $48.0\pm16.0$        & $26.0\pm8.6$          \\
Data                                & $10.00\pm0.50$                   & $49.7\pm1.7$         & $22.4\pm1.5$          \\ \hline
$B_s^0\rightarrow\phi \bar{K}^{*0}$ & $0.12_{-0.03-0.03-0.04}^{+0.04+0.04+0.06}$ & $74.5_{-4.8-3.7-6.4}^{+4.6+2.5+3.4}$ & $12.7_{-2.5-1.8-1.8}^{+2.4+2.2+3.3}$ \\
PQCD-I \cite{Ali:2007ff}            & $0.65^{+0.16+1.27+0.10}_{-0.13-0.18-0.04}$ & $71.2^{+3.2+2.7+0.0}_{-3.0-3.7-0.0}$ & $13.3^{+1.4+1.7+0.0}_{-1.5-1.3-0.0}$ \\
PQCD-II \cite{Zou:2015iwa}          & $0.39^{+0.20}_{-0.17}$           & $50.0^{+8.1}_{-7.2}$ & $24.2^{+3.6}_{-3.9}$  \\
QCDF-I \cite{Beneke:2006hg}         & $0.4^{+0.1+0.5}_{-0.1-0.3}$      & $40^{+1+67}_{-1-35}$ & $\cdots$              \\
QCDF-II \cite{Cheng:2009mu}         & $0.37^{+0.06+0.24}_{-0.05-0.20}$ & $43^{+2+21}_{-2-18}$ & $\cdots$              \\
QCDF-III~\cite{Chang:2017brr} \footnotemark[1] & $0.11^{+0.07+0.06}_{-0.04-0.01}$ & $43.6^{+14.6+51.5}_{-24.0-25.3}$  & $25.9^{+8.4+14.4}_{-9.1-23.5}$  \\
SCET \cite{Wang:2017rmh}            & $0.56\pm0.19$                    & $54.6\pm15.0$        & $20.5\pm9.1$          \\
FAT \cite{Wang:2017hxe}             & $0.70\pm0.11\pm0.13\pm0.08$      & $38.9\pm14.7$        & $31.4\pm8.1$          \\
Data                                & $1.14\pm0.30$                    & $51.0\pm17.0$        & $\cdots$              \\ \hline
$B^+\rightarrow\phi K^{*+}$         & $7.6_{-1.8-1.1-1.8}^{+2.9+1.4+2.8}$ & $72.3_{-3.9-2.6-0.5}^{+4.4+2.9+2.2}$ & $14.3_{-2.2-1.6-1.2}^{+1.9+1.2+0.1}$ \\
PQCD-I~\cite{Chen:2002pz}           & $15.96$                          & $74.8$               & $11.1$                \\
PQCD-II \cite{Zou:2015iwa}          & $10.3^{+4.9}_{-3.8}$             & $57.0^{+6.3}_{-5.9}$ & $21.0^{+3.0}_{-3.0}$  \\
QCDF-I \cite{Beneke:2006hg}         & $10.1^{+0.5+12.2}_{-0.5-7.1}$    & $45^{+0+58}_{-0-36}$ & $\cdots$              \\
QCDF-II \cite{Cheng:2008gxa}        & $10.0^{+1.4+12.3}_{-1.3-6.1}$    & $49^{+51}_{-42}$     & $25^{+21}_{-25}$      \\
SCET \cite{Wang:2017rmh}            & $9.86\pm3.39$                    & $51.0\pm16.4$        & $22.2\pm9.9$          \\
FAT \cite{Wang:2017hxe}             & $9.31\pm1.90\pm1.83\pm0.97$      & $48.0\pm16.0$        & $25.9\pm8.6$          \\
Data                                & $10.0\pm2.0$                     & $50.0\pm5.0$         & $20.0\pm5.0$          \\
\hline\hline
\end{tabular}
\footnotetext[1]{We quote the results of case II.}
\end{table}

Since the width-to-mass ratio for $\phi$ and $K^*(892)$ is small,
it is valid to apply the narrow width approximation for vector resonance to factorize the four-body process as three sequential two-body decays:
\begin{eqnarray}
\mathcal{B}(B \rightarrow \phi(\rightarrow K\bar K)K^*(\rightarrow K\pi))\approx
\mathcal{B}(B \rightarrow \phi K^*)\times \mathcal{B}(\phi \rightarrow K\bar K)\times \mathcal{B}(K^*\rightarrow K\pi),
\end{eqnarray}
for which we can extract the two-body $B \rightarrow \phi K^*$ branching ratios
 to compare with  the current available predictions and experiments.
 The longitudinal, perpendicular, and parallel polarization fractions of the $P$-wave  amplitudes are defined as
\begin{eqnarray}
f_0&=&\frac{\mathcal{B}_0}{\mathcal{B}_{P}},\quad
f_{\parallel}=\frac{\mathcal{B}_{\parallel}}{\mathcal{B}_{P}},\quad
f_{\perp}=\frac{\mathcal{B}_{\perp}}{\mathcal{B}_{P}},
\end{eqnarray}
with $\mathcal{B}_{P}=\mathcal{B}_0+\mathcal{B}_{\parallel}+\mathcal{B}_{\perp}$ being the total $P$-wave branching ratio.
The numerical results together with other theoretical results  from
PQCD \cite{Chen:2002pz, Ali:2007ff, Zou:2015iwa},
QCDF \cite{Beneke:2006hg, Cheng:2008gxa, Cheng:2009mu, Chang:2017brr},
SCET \cite{Wang:2017rmh} and FAT \cite{Wang:2017hxe}
are summarized in Table~\ref{tab:TwobodyBr} for comparison.
The world average values are taken from PDG~\cite{pdg2020} whenever available.

We see that the various approaches as well as experiment have similar branching ratios in magnitude for the $B^0$ and $B^+$ modes
but quite different results for $\mathcal{B}(B^0_s\rightarrow \phi \bar{K}^{*0})$.
The predicted central values span a wide range: $(1.1-7.0)\times 10^{-7}$,
 which are generally  below the current world average of $(1.14\pm0.30)\times 10^{-6}$.
Our result is consistent with the recent QCDF calculation~\cite{Chang:2017brr},
and closer to the predictions from Refs.~\cite{Cheng:2009mu,Beneke:2006hg,Zou:2015iwa},
but far from the previous PQCD value~\cite{Ali:2007ff}.
The discrepancy between theoretical predictions and experimental data remains an issue to be resolved.

According to the factorization assumption,
the polarization fractions for the vector-vector modes should satisfy the naive counting rules~\cite{Li:2004mp}
\begin{eqnarray}\label{eq:scale}
f_0\sim 1-\mathcal{O}(m_V^2/M^2),\quad f_{\parallel}\sim f_{\perp}\sim \mathcal{O}(m_V^2/M^2),
\end{eqnarray}
with $m_V$ being the vector meson mass.
The longitudinal polarization is naively expected to be $f_0\sim 0.9$ in $B\rightarrow\phi K^*$ decays.
However, a low longitudinal polarization of order $0.5$ has been  observed in the $B\rightarrow\phi K^*$ decays by Belle~\cite{Belle:2013vat,Belle:2005lvd},   $BABAR$~\cite{BaBar:2004uwv,BaBar:2008lan,BaBar:2006ttd}, and   LHCb~\cite{LHCb:2013nlu,LHCb:2014xzf},
which indicates a significant departure from the naive expectation of predominant longitudinal polarization
and poses  an interesting challenge for theoretical interpretations.
Several attempts to understand the values of  within or beyond the standard
model have been made~\cite{Grossman:2003qi,Alvarez:2004ci,Das:2004hq,Chen:2005mka,Yang:2004pm,Yang:2005tv,Baek:2005jk,Huang:2005qb,Chen:2006vs,Faessler:2007br,
Kagan:2004uw,Li:2004ti,Chen:2005cx,Beneke:2005we,Chen:2007qj,Datta:2007qb,
Beneke:2006hg,Cheng:2008gxa,Bauer:2004tj,Colangelo:2004rd,Ladisa:2004bp,Cheng:2004ru,Cheng:2010yd,Bobeth:2014rra}.

In the PQCD approach, a large transverse polarization fraction derives
from the weak annihilation diagram induced by the operator $O_6$ and nonfactorizable contributions~\cite{Li:2004ti}.
However, the combined  effects
are  not sufficient to reduce $f_0$ down to 0.5.
As can be seen from Table~\ref{tab:TwobodyBr}, our predictions for the longitudinal
polarization fractions are generally larger than $0.7$,
and agree with the previous PQCD calculations from Refs.~\cite{Chen:2002pz,Ali:2007ff}.
The small $f_0\sim (0.50-0.57)$ in Ref.~\cite{Zou:2015iwa} is ascribed to the
inclusion of the higher-power terms proportional to $r^2$,
with $r$ being the mass ratio between the vector and $B$ mesons.
The QCDF~\cite{Beneke:2006hg,Cheng:2008gxa} and FAT~\cite{Wang:2017hxe}
predictions on the longitudinal polarization fractions are generally less than 0.5.
A recent Belle measurement~\cite{Belle-II:2020rdz} based on the Summer 2020 Belle II dataset of 34.6 $fb^{-1}$,
yields $f_0(B^0\rightarrow \phi K^{*0})=0.57\pm0.20\pm0.04$ and $f_0(B^+\rightarrow \phi K^{*+})=0.58\pm0.23\pm0.02$,
to be compared with our results.

\subsection{$CP$-violating observables}
\begin{table}[!htbh]
\caption{  Direct $CP$ asymmetries (in units of $\%$) for
the four-body  $B^+\rightarrow(K^+K^-)(K^0\pi^+)$ decay.
The requirements on the $K\bar K$ and $K\pi$ invariant masses
are $m_{\phi}-0.015<m_{K\bar K}<m_{\phi}+0.015$ (GeV) and $m_{K^*}-0.15<m_{K\pi}<m_{K^*}+0.15$ (GeV).
The sources of theoretical errors are the same
as in previous tables but added in quadrature.}
\label{tab:DCPV}
\begin{tabular}[t]{lcccccc}
\hline\hline
 $\mathcal{A}_0^{CP}$ & $\mathcal{A}^{CP}_\parallel$ & $\mathcal{A}^{CP}_\perp$ & $\mathcal{A}^{CP}_{SS}$ & $\mathcal{A}^{CP}_{VS}$ & $\mathcal{A}^{CP}_{SV}$ & $\mathcal{A}^{CP}_{\text{total}}$ \\ \hline
 $-4.1_{-4.6}^{+6.1}$ & $5.8_{-3.8}^{+2.1}$          & $4.9_{-4.4}^{+3.8}$      & $4.5_{-2.5}^{+2.0}$     & $3.8_{-4.2}^{+2.2}$     & $3.2_{-4.4}^{+0.1}$     & $-0.3_{-2.5}^{+3.2}$ \\
\hline\hline
\end{tabular}
\end{table}

\begin{table}[!htbh]
\caption{Theoretical predictions of $CP$ violation (in \%) for the $B^+\rightarrow\phi K^{*+}$ decay in various approaches.}
\label{tab:DCPVpk}
\begin{tabular}[t]{lcccccc}
\hline\hline
This Work & Data \cite{pdg2020} & PQCD \cite{Zou:2015iwa}& QCDF \cite{Beneke:2006hg}& QCDF \cite{Cheng:2009cn} &SCET \cite{Wang:2017rmh} & FAT \cite{Wang:2017hxe} \\ \hline
$-1.5_{-3.3}^{+4.9}$ & $-1\pm8$ & $-1.0$ & $0^{+0+2}_{-0-1}$        &0.05                      & $-0.39\pm0.44$          & $1.00\pm0.27$           \\
\hline\hline
\end{tabular}
\end{table}

The direct $CP$ asymmetry in each component and the overall asymmetry are defined as
\begin{eqnarray}
\mathcal{A}_h^{CP}=\frac{\mathcal{\bar{B}}_h-\mathcal{B}_h}{\mathcal{\bar{B}}_h+\mathcal{B}_h},\quad
\mathcal{A}_{\text{total}}^{CP}=\frac{\sum_h \mathcal{\bar{B}}_h-\sum_h \mathcal{B}_h}{\sum_h \mathcal{\bar{B}}_h+\sum_h \mathcal{B}_h},
\end{eqnarray}
respectively.
Since only penguin operators work on the neutral channels, there is no direct $CP$ asymmetry in the neutral $B^0$ and $B^0_s$ modes.
However, the charged mode receives an additional tree contribution
and the direct $CP$ asymmetries arises from the interference between the tree and penguin amplitudes.
As shown in Table~\ref{tab:DCPV},
the direct $CP$ asymmetries for various helicity states turn out to be small, $\sim \mathcal{O}(10^{-2})$,
and the overall $CP$ asymmetry is even lower at the order of $10^{-3}$.
It can be understood as follows.
 The tree contribution only appears in the annihilation diagrams,
 which are power suppressed  with respect to the emission ones.
Furthermore, the CKM element $|V^*_{ub}V_{us}|$ of tree diagrams is smaller than $|V^*_{tb}V_{ts}|$ of penguin diagrams.
Our result in Table~\ref{tab:DCPV} for the $VS$ and $SV$ components are
consistent with the world averages of $(4\pm 16)\%$ and $(-15\pm 12)\%$~\cite{pdg2020}, respectively, within uncertainties.

In Table~\ref{tab:DCPVpk}, we list the predicted direct $CP$ asymmetry of the mode $B^+\rightarrow \phi K^{*+}$
in PQCD. For comparison, the experimental data, as well as
predictions from previous PQCD~\cite{Zou:2015iwa}, QCDF~\cite{Beneke:2006hg,Cheng:2009cn}, SCET~\cite{Wang:2017rmh},
and FAT~\cite{Wang:2017hxe}, are also presented.
All the theoretical approaches show that
a nearly vanishing direct $CP$ asymmetry,
complies with the latest word average of $-0.01\pm0.08$~\cite{pdg2020} from the measurements~\cite{BaBar:2007bpi,Belle:2005lvd}
\begin{eqnarray}
\mathcal{A}^{CP}(B^+\rightarrow \phi K^{*+})=\left\{
\begin{aligned}
&0.00\pm0.09(\text{stat})\pm0.04(\text{syst}) \quad\quad\quad  &\text{$BABAR$}, \nonumber\\ 
&-0.02 \pm 0.14 (\text{stat})\pm 0.03 (\text{syst}) \quad\quad\quad  & \text{Belle}.  \nonumber\\ 
\end{aligned}\right.
\end{eqnarray}
Any observation of large direct $CP$ asymmetry to this mode
will be a signal for new physics.

TPAs, as mentioned in the Introduction, may be potential signals of $CP$ violation,
and thus are complementary to the direct $CP$ violations,
particularly when the latter are suppressed by the strong phase.
According to Eq.~(\ref{eq:at}),
 TPAs can be calculated from integrations of the differential decay rate as
\begin{eqnarray}\label{eq:TPAs}
A_T^1&=&\frac{\Gamma((2\zeta_1-1)(2\zeta_2-1)\sin \varphi  >0)-\Gamma((2\zeta_1-1)(2\zeta_2-1)\sin\varphi <0)}
{\Gamma((2\zeta_1-1)(2\zeta_2-1)\sin\varphi >0)+\Gamma((2\zeta_1-1)(2\zeta_2-1)\sin\varphi <0)} \nonumber\\
&=&-\frac{2\sqrt{2}}{\pi\mathcal{D}}\int d\omega_1d\omega_2k(\omega_1)k(\omega_2)k(\omega_1,\omega_2)Im[ A_\perp A_0^*], \nonumber\\
A_T^2&=&\frac{\Gamma(\sin(2\varphi) >0) - \Gamma(\sin(2\varphi) <0)}{\Gamma(\sin(2\varphi) >0) + \Gamma(\sin(2\varphi) <0)} \nonumber\\
&=&-\frac{4}{\pi\mathcal{D}}\int d\omega_1d\omega_2k(\omega_1)k(\omega_2)k(\omega_1,\omega_2)Im[A_\perp A_\parallel^*], \nonumber\\
A_T^3&=&\frac{\Gamma((2\zeta_1-1)\sin\varphi >0)-\Gamma((2\zeta_1-1)\sin\varphi <0)}
{\Gamma((2\zeta_1-1)\sin\varphi >0)+\Gamma((2\zeta_1-1)\sin\varphi <0)} \nonumber\\
&=&-\frac{3}{\sqrt{2}\mathcal{D}}\int d\omega_1d\omega_2k(\omega_1)k(\omega_2)k(\omega_1,\omega_2)Im[A_\perp A_{VS}^*], \nonumber\\
A_T^4&=&\frac{\Gamma((2\zeta_2-1)\sin\varphi >0)-\Gamma((2\zeta_2-1)\sin\varphi <0)}
{\Gamma((2\zeta_2-1)\sin\varphi >0)+\Gamma((2\zeta_2-1)\sin\varphi <0)} \nonumber\\
&=&-\frac{3}{\sqrt{2}\mathcal{D}}\int d\omega_1d\omega_2k(\omega_1)k(\omega_2)k(\omega_1,\omega_2)Im[A_\perp A_{SV}^*], \nonumber\\
A_T^5&=&\frac{\Gamma(\sin\varphi >0)-\Gamma(\sin\varphi <0)}{\Gamma(\sin\varphi >0)+\Gamma(\sin\varphi <0)} \nonumber\\
&=&-\frac{9\pi}{4\sqrt{2}\mathcal{D}}\int d\omega_1d\omega_2k(\omega_1)k(\omega_2)k(\omega_1,\omega_2)Im[A_\perp A_{SS}^*],
\end{eqnarray}
with
\begin{eqnarray}
\mathcal{D}=\int d\omega_1d\omega_2k(\omega_1)k(\omega_2)k(\omega_1,\omega_2)\sum_hY_h|A_h|^2, 
\end{eqnarray}
where the mass integration extends over the chosen  mass window.
It is seen that the TPAs are produced by the interference between $A_{\perp}$ and $A_{i}$ with $i=0,\parallel,SV,VS,SS$ and
take the form  $Im(A_{\perp}A_i^*)=|A_{\perp}||A_i^*|\sin(\Delta\phi+\Delta \delta)$,
where $\Delta\phi$ and $\Delta\delta$, respectively, denote weak and strong phase differences between the two amplitudes.
Here the strong phase difference could produce a nonzero value, even if the weak phases vanish.
Thus, a nonzero TPA is not necessarily a signal of $CP$ violation.
In order to obtain a true signal of $CP$ violation, one has to compare the TPAs in  $B$ and $\bar{B}$ decays.
The helicity amplitudes for the $CP$-conjugated process can be obtained  by applying the following transformations:
\begin{eqnarray}
A_0\rightarrow \bar{A}_0, A_\parallel\rightarrow \bar{A}_\parallel, A_\perp\rightarrow -\bar{A}_\perp,
 A_{SV}\rightarrow \bar{A}_{SV}, A_{VS}\rightarrow \bar{A}_{VS},A_{SS}\rightarrow \bar{A}_{SS}.
\end{eqnarray}
Then, the associated TPAs for the charge-conjugate process, $\bar{A}_T^i$,  are defined similarly. 
Now, we can construct the true and fake asymmetries
by combining $A_T^i$ and $\bar{A}_T^i$~\cite{Belle-II:2018jsg}
\begin{eqnarray}\label{eq:ture1}
A_T^i(true)&=&\frac{1}{2}(A_T^i+\bar{A}_T^i)\propto \sin(\Delta\phi)\cos(\Delta \delta),\nonumber\\
A_T^i(fake)&=&\frac{1}{2}(A_T^i-\bar{A}_T^i)\propto \cos(\Delta\phi)\sin(\Delta \delta).
\end{eqnarray}
It was pointed out in Ref.~\cite{Gronau:2011cf} that the second equation above is valid
only in the absence of direct $CP$ asymmetry in the total decay rate.
As the total direct $CP$ asymmetry does not exceed a few percent as shown in Table~\ref{tab:DCPV},
the above approximation holds in $B\rightarrow \phi K^*$ decays.
It is clear that $A_T^i(true)$ do not suffer the suppression from the strong phase $\Delta \delta$ compared with the direct $CP$ violation.
It is nonzero only if the weak phases are nonzero, and provides a measure for $CP$ violation.
Nevertheless,
$A_T^i(fake)$ can be nonzero even if the weak phases are zero.
Such a quantity will sometimes be referred to as a fake asymmetry.
It reflects the importance of strong final-state phases~\cite{Gronau:2011cf},
and thus is not a signal of $CP$ violation.
Since the helicity amplitudes always have different strong phases,
this will lead to nonzero fake TPAs for all decays.

\begin{table}[!htbh]
\caption{PQCD predictions for the TPAs(\%).
The mass of the $K\bar K(K\pi)$ pair is required to be within
15 MeV (150 MeV) of the known $\phi(K^*)$ meson mass.}
\label{tab:TPAs}
\begin{tabular}[t]{lccc}
\hline\hline
Asymmetries    & $B^0\rightarrow(K^+K^-)(K^+\pi^-)$ & $B_s^0\rightarrow(K^+K^-)(K^-\pi^+)$ & $B^+\rightarrow(K^+K^-)(K^0\pi^+)$  \\ \hline
$A^1_T$       & $-13.8^{+4.8}_{-4.3}$               & $-24.0^{+6.6}_{-3.8}$                 & $-14.1^{+5.0}_{-3.8}$                \\
$\bar{A}^1_T$ & $13.8^{+4.8}_{-4.3}$              & $24.0^{+6.6}_{-3.8}$                & $+13.8^{+5.6}_{-3.9}$               \\
$A^1_T(true)$ & $0.0$                              & $0.0$                                & $-0.15^{+0.05}_{-0.30}$              \\
$A^1_T(fake)$ & $-13.8^{+4.8}_{-4.3}$               & $-24.0^{+6.6}_{-3.8}$                 & $-14.0^{+5.3}_{-3.9}$                \\ \hline
$A^2_T$       & $-0.3_{-0.1}^{+0.1}$               & $-0.1_{-0.0}^{+0.0}$                 & $-0.3_{-0.1}^{+0.1}$                \\
$\bar{A}^2_T$ & $0.3_{-0.1}^{+0.1}$                & $0.1_{-0.0}^{+0.0}$                  & $0.2_{-0.0}^{+0.1}$                 \\
$A^2_T(true)$ & $0.0$                              & $0.0$                                & $-0.05_{-0.05}^{+0.00}$             \\
$A^2_T(fake)$ & $-0.3_{-0.1}^{+0.1}$               & $-0.1_{-0.0}^{+0.0}$                 & $-0.3_{-0.1}^{+0.1}$                \\ \hline
$A^3_T$       & $-5.4_{-0.6}^{+1.0}$               & $-6.4_{-2.2}^{+2.1}$                 & $-5.6_{-0.6}^{+1.0}$                \\
$\bar{A}^3_T$ & $5.4_{-0.6}^{+1.0}$                & $6.4_{-2.2}^{+2.1}$                  & $5.5_{-0.6}^{+1.0}$                 \\
$A^3_T(true)$ & $0.0$                              & $0.0$                                & $-0.05_{-0.00}^{+0.00}$             \\
$A^3_T(fake)$ & $-5.4_{-0.6}^{+1.0}$               & $-6.4_{-2.2}^{+2.1}$                 & $-5.6_{-0.6}^{+1.0}$                \\ \hline
$A^4_T$       & $1.6_{-3.0}^{+3.0}$                & $-8.1_{-3.3}^{+3.2}$                 & $1.6_{-2.7}^{+3.2}$                 \\
$\bar{A}^4_T$ & $-1.6_{-3.0}^{+3.0}$               & $8.1_{-3.3}^{+3.2}$                  & $-1.8_{-2.8}^{+3.1}$                \\
$A^4_T(true)$ & $0.0$                              & $0.0$                                & $-0.10_{-0.00}^{+0.05}$             \\
$A^4_T(fake)$ & $1.6_{-3.0}^{+3.0}$                & $-8.1_{-3.3}^{+3.2}$                 & $1.7_{-2.8}^{+3.2}$                 \\ \hline
$A^5_T$       & $-4.3_{-0.8}^{+1.1}$               & $-6.7_{-1.9}^{+1.8}$                 & $-4.2_{-0.8}^{+1.1}$                \\
$\bar{A}^5_T$ & $4.3_{-0.8}^{+1.1}$                & $6.7_{-1.9}^{+1.8}$                  & $4.3_{-0.8}^{+1.2}$                 \\
$A^5_T(true)$ & $0.0$                              & $0.0$                                & $0.05_{-0.05}^{+0.00}$              \\
$A^5_T(fake)$ & $-4.3_{-0.8}^{+1.1}$               & $-6.7_{-1.9}^{+1.8}$                 & $-4.3_{-0.8}^{+1.2}$                \\
\hline\hline
\end{tabular}
\end{table}

The calculated TPAs for the concerned decays are collected in Table~\ref{tab:TPAs}.
In the special case of the involved neutral intermediate states $B^0\rightarrow \phi K^{*0}$ and $B^0_s\rightarrow \phi \bar{K}^{*0}$ modes,
in which each helicity amplitude involves the same  single weak phase in the SM.
This results in $A_T^i=-\bar{A}_T^i$  due to the vanishing weak phase difference.
The true TPAs for the two neutral modes  are thus expected to be zero as shown in Table~\ref{tab:TPAs}. 
If such asymmetries are observed experimentally, it is probably a signal of new physics.
The situation  of the fake  TPAs is different. Many nonzero and even sizable fake TPAs are predicted in our calculations.
The magnitude of $A_T^1(fake)$ for the $B^0$ and $B^+$ channels exceeds ten percent,
even reaching $24.0\%$ for the $B_s$ one,
whereas the $S$-wave induced fake TPAs, $A_T^{3,4,5}(fake)$, are predicted to be several percent.
The smallness of $A_T^2(fake)$ is caused by the suppression from the strong phase difference
 between the perpendicular and parallel polarization amplitudes,
which are found  to be very close to 0 in  PQCD framework~\cite{Zou:2015iwa,Ali:2007ff}.
Hence, the measurement of a large $A_T^2$  would point clearly towards the presence of new physics beyond the SM.

\begin{table}[!htbh]
\caption{Comparison of measurements in the angular  analysis of $B^0\rightarrow \phi K^{*0}$ made by  $BABAR$~\cite{BaBar:2008lan},
Belle~\cite{Belle:2013vat} and LHCb~\cite{LHCb:2014xzf} experiments,
where the first and second uncertainties are statistical and systematic, respectively.
The true and fake TPAs for the $P$-wave components are deduced from the fitted parameters (the first eight lines),
while the $S$-wave ones are taken from LHCb~\cite{LHCb:2014xzf}.}
\label{tab:TPAsdata}
\begin{tabular}[t]{lccc}
\hline\hline
Parameters               & $BABAR$                    & Belle                    & LHCb                     \\ \hline
$f_L$                   & $0.494\pm0.034\pm0.013$  & $0.499\pm0.030\pm0.018$  & $0.497\pm0.019\pm0.015$  \\
$f_\perp$               & $0.212\pm0.032\pm0.013$  & $0.238\pm0.026\pm0.008$  & $0.221\pm0.016\pm0.013$  \\
$\delta_\perp$          & $2.35\pm0.13\pm0.09$     & $2.37\pm0.10\pm0.04$     & $2.633\pm0.062\pm0.037$  \\
$\delta_\parallel$      & $2.40\pm0.13\pm0.08$     & $2.23\pm0.10\pm0.02$     & $2.562\pm0.069\pm0.040$  \\
$A_0^{CP}$              & $+0.01\pm0.07\pm0.02$    & $-0.030\pm0.061\pm0.007$ & $-0.003\pm0.038\pm0.005$ \\
$A_\perp^{CP}$          & $-0.04\pm0.15\pm0.06$    & $-0.14\pm0.11\pm0.01$    & $+0.047\pm0.072\pm0.009$ \\
$\delta_\perp^{CP}$     & $+0.21\pm0.13\pm0.08$    & $+0.05\pm0.10\pm0.02$    & $+0.062\pm0.062\pm0.006$ \\
$\delta_\parallel^{CP}$ & $+0.22\pm0.12\pm0.08$    & $-0.02\pm0.10\pm0.01$    & $+0.045\pm0.068\pm0.015$ \\ \hline
Asymmetries              & $BABAR$                    & Belle                    & LHCb                     \\ \hline
$A^1_T(true)$           & $-0.046\pm0.031\pm0.017$ & $-0.029\pm0.025\pm0.005$ & $-0.007\pm0.012\pm0.002$ \\
$A^2_T(true)$           & $-0.003\pm0.056\pm0.036$ & $0.021\pm0.040\pm0.006$  & $+0.004\pm0.014\pm0.002$ \\
$A^3_T(true)$           & $\cdots$                 & $\cdots$                 & $+0.004\pm0.006\pm0.001$ \\
$A^4_T(true)$           & $\cdots$                 & $\cdots$                 & $+0.002\pm0.006\pm0.001$ \\
$A^1_T(fake)$           & $-0.203\pm0.031\pm0.019$ & $-0.211\pm0.025\pm0.010$ & $-0.105\pm0.012\pm0.006$ \\
$A^2_T(fake)$           & $0.016\pm0.058\pm0.038$  & $-0.041\pm0.040\pm0.013$ & $-0.017\pm0.014\pm0.003$ \\
$A^3_T(fake)$           & $\cdots$                 & $\cdots$                 & $-0.063\pm0.006\pm0.005$ \\
$A^4_T(fake)$           & $\cdots$                 & $\cdots$                 & $-0.019\pm0.006\pm0.007$ \\
\hline\hline
\end{tabular}
\end{table}
\begin{table}[!htbh]
\caption{Parameters measured in the angular  analysis of $B^+\rightarrow \phi K^{*+}$  by $BABAR$~\cite{BaBar:2007bpi},
 where the first and second uncertainties are statistical and systematic, respectively.
 The true and fake TPAs (the last four entries) are deduced from the fitted parameters (the first eight lines).}
\label{tab:TPAsdata1}
\begin{tabular}[t]{lc}
\hline\hline
Parameters               & $BABAR$                   \\ \hline
$f_L$                   & $+0.49\pm0.05\pm0.03$    \\
$f_\perp$               & $+0.21\pm0.05\pm0.02$    \\
$\delta_\perp-\pi$      & $-0.45\pm0.20\pm0.03$    \\
$\delta_\parallel-\pi$  & $-0.67\pm0.20\pm0.07$    \\
$A_0^{CP}$              & $+0.17\pm0.11\pm0.02$    \\
$A_\perp^{CP}$          & $+0.22\pm0.24\pm0.08$    \\
$\delta_\perp^{CP}$     & $+0.19\pm0.20\pm0.07$    \\
$\delta_\parallel^{CP}$ & $+0.07\pm0.20\pm0.05$    \\ \hline
Asymmetries              & $BABAR$                    \\ \hline
$A^1_T(true)$           & $-0.025\pm0.056\pm0.019$ \\
$A^2_T(true)$           & $0.028\pm0.084\pm0.026$  \\
$A^1_T(fake)$           & $-0.114\pm0.056\pm0.011$ \\
$A^2_T(fake)$           & $-0.061\pm0.084\pm0.023$ \\
\hline\hline
\end{tabular}
\end{table}

Experimentally, an complete angular analysis of the decay $B^0\rightarrow \phi K^{*0}$  is available from  LHCb~\cite{LHCb:2014xzf},
 $BABAR$~\cite{BaBar:2008lan,BaBar:2006ttd}, and Belle~\cite{Belle:2013vat,Belle:2005lvd} Collaborations,
which allows us  to determine the true and fake TPAs from the measured polarization amplitudes, phases, and amplitude differences between
$B^0$ and $\bar B^0$ decays.
The concerned decays are all self-tagged processes,
whose triple product asymmetry can be computed separately for $B$ and $\bar B$ decays.
We note that the definitions of the $CP$ asymmetries and TPAs are different among these measurements.
To compare these results directly, it is  necessary to rescale them with the unified definitions.
We follow the convention of Ref.~\cite{Gronau:2011cf},
defining
\begin{eqnarray}
(A_T^1)_{\text{exp}}&=&-\frac{2\sqrt{2}}{\pi}\frac{Im(A_{\perp}A_0^*)}{|A_0|^2+|A_\parallel|^2+|A_{\perp}|^2},\nonumber\\
(A_T^2)_{\text{exp}}&=&-\frac{4}{\pi}\frac{Im(A_{\perp}A_\parallel^*)}{|A_0|^2+|A_\parallel|^2+|A_{\perp}|^2}.
\end{eqnarray}
The corresponding quantities for the charge-conjugate process, $(\bar A_T^{1,2})_{\text{exp}}$, are defined similarly.
We can calculate the true and fake asymmetries  from Eq.~(\ref{eq:ture1}).
The measured polarizations, phases, and $CP$ violation asymmetries as well as the true and fake TPAs of $B^0\rightarrow \phi K^{*0}$
 by the LHCb, $BABAR$, and Belle Collaborations are compared in Table~\ref{tab:TPAsdata}.
One can  see that  all the true asymmetries are measured to be consistent with zero,
showing no evidence for $CP$ violation.
Nevertheless, a significant fake asymmetry, such as $A_T^1(fake)$, is observed in all three different experiments,
reflecting the importance of final-state interactions.
The $S$-wave induced TPAs, which arise from the interference between $A_\perp$ and one $S$-wave amplitude,
have not been determined in $BABAR$ and Belle,
since the contributions of $S$-wave $K\bar K$ and
$K\pi$ and their interferences  were not fully included into the angular analysis.
It is pointed out that contributions from the $S$-wave $K\bar K$ and $K\pi$ are significant
and should be taken into account in future measurements.
The three additional $CP$-violating observables can then provide valuable complementary information on NP.

The only available amplitude analysis of $B^+\rightarrow \phi K^{*+}$  was performed by the $BABAR$~\cite{BaBar:2007bpi} experiment.
The fitted polarization parameters together with  the true and fake TPAs are summarized in Table~\ref{tab:TPAsdata1}.
The $S$-wave $K\pi$ contribution was included to resolve the twofold phase ambiguity,
but the accurate assessments of the $S$-wave component have not been determined,
leading to the corresponding $S$-wave induced TPAs still being absent.

Although the polarization fractions and branching ratio of the decay $B^0_s \rightarrow \phi \bar{K}^{*0}$
have been measured by the LHCb experiment~\cite{LHCb:2013nlu},
the full angular analysis has not been done because of the limited signal events, 
which results in the measurement on TPAs not being available yet.
As indicated in Table~\ref{tab:TPAs},
the predicted large fake TPAs for the $B_s$ mode would be tested in the future.
\section{conclusion}\label{sec:sum}
$B$ meson four-body decays provide a wealth of information on the weak interactions,
in terms of a number of observables ranging from the branching ratios, polarizations,
$CP$ asymmetries, and triple product asymmetries to a full angular analysis.
In this work, we have concentrated on the penguin-dominated $B_{(s)}\rightarrow \phi(\rightarrow K\bar K)K^*(\rightarrow K\pi)$ decays
using the perturbative QCD approach.
The calculations were performed in the $K\bar K(K\pi)$ invariant mass window of 15(150) MeV around the $\phi (K^*)$ mass.
In addition to the dominant vector resonances,
four-particle final states can also be obtained through one or two scalar meson intermediate states in the given mass regions.
The strong dynamics of the scalar or vector resonance  decays into the meson pair
was parametrized into the corresponding  two-meson DAs,
which has been well established in three-body $B$ meson decays.
Angular momentum conservation in the decay allows for three amplitudes for vector-vector decays
and one single amplitude in modes involving at least one scalar $K\bar K (K\pi)$ pair.
The $CP$-averaged branching ratios of all components were predicted in the chosen mass ranges.
The single $S$-wave contributions were found to be significant and
 consistent with the data from the LHCb experiment,
 while the double $S$-wave contributions were only a few percent in the considered invariant mass range.
It has been demonstrated that the $S$-wave contributions were strongly sensitive to the integrating ranges.
After choosing the same $K\pi$ mass range, our results can be comparable with the Belle and $BABAR$ measurements roughly.

We extracted the two-body $B\rightarrow \phi K^*$ branching ratios from the results for the
corresponding quasi-two-body modes by employing the narrow width approximation.
The obtained results  basically agree with previous predictions performed in the two-body framework within theoretical uncertainty.
However, various predictions for $\mathcal{B}(B_s^0\rightarrow \phi \bar K^{*0})$ lie in a wide range,
generally below the current world average value.
The gap between theory and experiment requires a more thorough study.
The longitudinal polarization fractions were estimated to be  $f_0 \sim 0.7$,
somewhat lower than the naive expectation of a dominant longitudinal polarization
because of the important chirally enhanced annihilation and  nonfactorizable contributions.
However, the observation of  an even smaller value of 0.5 by Belle, $BABAR$, and LHCb,
means  the existing explanations of the abnormal polarization are not satisfactory and thus call for more in-depth studies.

We have also investigated the direct $CP$ asymmetries and TPAs in the $B\rightarrow ( K\bar K)(K\pi)$ decays.
For the two pure-penguin neutral channels, both the direct $CP$ asymmetries and true TPAs should be zero in SM
due to the vanishing weak phase difference.
It was observed that the direct $CP$ asymmetries for the charged modes are small, of order $10^{-2}$,
since the tree contributions are significantly suppressed compared to the penguin ones.
A similar observation was made in previous PQCD, QCDF, SCET, and FAT approaches and supported by the measurements from $BABAR$ and Belle.
The true TPAs are predicted to be tiny, of order $10^{-3}$, compatible with the absence of $CP$ violation,
whereas, the fake TPAs  were found to be sizable, which can provide valuable information on final-state interactions.

The full angular analysis of $B^0\rightarrow \phi K^{*0}$ has been performed by LHCb, $BABAR$, and Belle experiments,
and allow one to derive the true and fake TPAs from the results of the angular analysis.
It is worth noting that the $S$-wave components and their interference have not been
 fully considered into the analysis in $BABAR$ and Belle experiments.
Therefore, the only available measurements for the $S$-wave induced TPAs are from LHCb.
By including the $S$-wave $K\bar K$ and $K\pi$ components, we have estimated the $S$-wave induced TPAs for the first time.
The predicted asymmetries $B^0\rightarrow \phi K^{*0}$ are in good agreement with those reported by the LHCb.
As the three additional $CP$-violating observables may provide valuable complementary information on NP,
a dedicated angular analysis from Belle and $BABAR$ data is expected.
The full  angular analysis of $B^0_s\rightarrow \phi \bar K^{*0}$ was still not available due to the limited samples.
The obtained asymmetries can be confronted with the future data.

The experimental measurement may be improved by expanding the mass region to around $1.5$ GeV based on  more precise data in the future.
Then the angular analysis could include the contributions of some higher excited intermediate states,
such as $D$-wave $K^*_2(1430)$, $P$-wave $\phi(1680)$, and $S$-wave $f_0(1370)$, and so on.
The additional contributions and their interferences would provide a lot of  meaningful asymmetries in angular distributions,
and leading to the amplitude analysis will be more complicated and complete.
This is an intriguing topic for future investigation.

\begin{acknowledgments}
This work is supported by National Natural Science Foundation of China under Grants No.~12075086, No.~12005103, and No.~11605060.
Z.R. is supported in part by the Natural Science Foundation of Hebei Province under Grants No.~A2021209002 and No.~A2019209449.
Y.L. is also supported by the Natural Science Foundation of Jiangsu Province under
Grant No.~BK20190508 and the Research Start-up Funds of Nanjing Agricultural University.
\end{acknowledgments}

\begin{appendix}
\section{Decay Amplitudes}\label{sec:ampulitude}
Here we present the helicity amplitudes for the concerned channels as follows:
\begin{eqnarray}
\mathcal{A}_h(B^0 \rightarrow \phi(\rightarrow K^+K^-) K^{*0}(\rightarrow K^+\pi^-))
&=&\frac{G_F}{\sqrt{2}}V^*_{tb}V_{ts}X_h\Big\{
\frac{4}{3}\Big [C_3+C_4-\frac{1}{2}(C_9+C_{10})\Big ]\mathcal{F}_e^{LL}\nonumber\\&&
+\Big [C_5+\frac{1}{3}C_6-\frac{1}{2}(C_7+\frac{1}{3}C_8)\Big ]\mathcal{F}_e^{LR}\nonumber\\&&
+\Big [C_6+\frac{1}{3}C_5-\frac{1}{2}(C_8+\frac{1}{3}C_{7})\Big ]\mathcal{F}_e^{SP}\nonumber\\&&
+\Big [C_3+C_4-\frac{1}{2}(C_9+C_{10})\Big ]\mathcal{M}_e^{LL}\nonumber\\&&
+(C_5-\frac{1}{2}C_7)\mathcal{M}_e^{LR}+(C_6-\frac{1}{2}C_8)\mathcal{M}_e^{SP}\nonumber\\&&
+\Big [C_4+\frac{1}{3}C_3-\frac{1}{2}(C_{10}+\frac{1}{3}C_9)\Big ]\mathcal{F}_a^{LL}\nonumber\\&&
+\Big [C_6+\frac{1}{3}C_5-\frac{1}{2}(C_8+\frac{1}{3}C_7)\Big ]\mathcal{F}_a^{SP}\nonumber\\&&
+(C_3-\frac{1}{2}C_9)\mathcal{M}_a^{LL}+(C_5-\frac{1}{2}C_7)\mathcal{M}_a^{LR}\Big\},
\end{eqnarray}

\begin{eqnarray}
\mathcal{A}_h(B_s^0 \rightarrow \phi(\rightarrow K^+ K^-) \bar{K}^{*0}(\rightarrow K^-\pi^+))
&=&\frac{G_F}{\sqrt{2}}V^*_{tb}V_{td}X_h\Big\{
\Big [C_3+\frac{1}{3}C_4-\frac{1}{2}(C_9+\frac{1}{3}C_{10})\Big ]\mathcal{F}_e^{LL}\nonumber\\&&
+\Big [C_4+\frac{1}{3}C_3-\frac{1}{2}(C_{10}+\frac{1}{3}C_{9})\Big ]\mathcal{F'}_e^{LL}\nonumber\\&&
+\Big [C_6+\frac{1}{3}C_5-\frac{1}{2}(C_8+\frac{1}{3}C_{7})\Big ]\mathcal{F'}_e^{SP}\nonumber\\&&
+\Big [C_5+\frac{1}{3}C_6-\frac{1}{2}(C_7+\frac{1}{3}C_{8})\Big ]\mathcal{F}_e^{LR}\nonumber\\&&
+(C_4-\frac{1}{2}C_{10})\mathcal{M}_e^{LL}+(C_3-\frac{1}{2}C_9)\mathcal{M'}_e^{LL}\nonumber\\&&
+(C_6-\frac{1}{2}C_{8})\mathcal{M}_e^{SP}+(C_5-\frac{1}{2}C_7)\mathcal{M'}_e^{LR}\nonumber\\&&
+\Big [C_4+\frac{1}{3}C_3-\frac{1}{2}(C_{10}+\frac{1}{3}C_{9})\Big ]\mathcal{F'}_a^{LL}\nonumber\\&&
+\Big [C_6+\frac{1}{3}C_5-\frac{1}{2}(C_8+\frac{1}{3}C_{7})\Big ]\mathcal{F'}_a^{SP}\nonumber\\&&
+(C_3-\frac{1}{2}C_9)\mathcal{M'}_a^{LL}+(C_5-\frac{1}{2}C_7)\mathcal{M'}_a^{LR}\Big\},
\end{eqnarray}

\begin{eqnarray}
\mathcal{A}_h(B^+ \rightarrow \phi(\rightarrow K^+K^-) K^{*+}(\rightarrow K^0\pi^+))
&=&\frac{G_F}{\sqrt{2}}V^*_{ub}V_{us}X_h\Big\{(C_2+\frac{1}{3}C_1)\mathcal{F}_a^{LL}+(C_1)\mathcal{M}_a^{LL}\Big\}\nonumber\\&&
-\frac{G_F}{\sqrt{2}}V^*_{tb}V_{ts}X_h\Big\{\frac{4}{3}\Big [C_3+C_4-\frac{1}{2}(C_9+C_{10})\Big ]\mathcal{F}_e^{LL}\nonumber\\&&
+\Big [C_5+\frac{1}{3}C_6-\frac{1}{2}(C_7+\frac{1}{3}C_{8})\Big ]\mathcal{F}_e^{LR}\nonumber\\&&
+\Big [C_6+\frac{1}{3}C_5-\frac{1}{2}(C_8+\frac{1}{3}C_{7})\Big ]\mathcal{F}_e^{SP}\nonumber\\&&
+\Big [C_3+C_4-\frac{1}{2}(C_9+C_{10})\Big ]\mathcal{M}_e^{LL}\nonumber\\&&
+(C_5-\frac{1}{2}C_7)\mathcal{M}_e^{LR}+(C_6-\frac{1}{2}C_{8})\mathcal{M}_e^{SP}\nonumber\\&&
+\Big [C_4+\frac{1}{3}C_3+C_{10}+\frac{1}{3}C_{9}\Big ]\mathcal{F}_a^{LL}\nonumber\\&&
+\Big [C_6+\frac{1}{3}C_5+C_8+\frac{1}{3}C_{7}\Big ]\mathcal{F}_a^{SP}\nonumber\\&&
+(C_3+C_9)\mathcal{M}_a^{LL}+(C_5+C_7)\mathcal{M}_a^{LR}\Big\},
\end{eqnarray}
with
\begin{eqnarray}
X_h = \begin{cases}
\sqrt{1+4\alpha_1}\sqrt{1+4\alpha_2}, & h=0, \parallel, \perp \\
\sqrt{1+4\alpha_{1,2}},               & h=SV, VS \\
1,                                    & h=SS.
\end{cases}
\end{eqnarray}
The explicit expressions of $\mathcal{F/M}$ can be found in~\cite{zjhep}.
Note that the term $\mathcal{F}_e^{SP}$ from the operators $O_{5-8}$ vanishes when a vector resonance is emitted from the weak vertex,
because neither the scalar nor the pseudoscalar density gives contributions to the vector resonance production.

\end{appendix}

\end{document}